\documentclass[final]{IEEEtran}
\usepackage[noadjust]{cite}
\usepackage{xcolor,soul,framed} 
\usepackage[utf8]{inputenc}
\usepackage[hidelinks]{hyperref} 
\usepackage{amsmath}
\usepackage{bm} 
\usepackage{graphics}
\usepackage{epsfig}
\usepackage{pdfpages}
\usepackage{statmath}
\usepackage{amssymb}
\usepackage{multirow}

\begin{document}
\bstctlcite{IEEEexample:BSTcontrol}

\title{A Novel Collaborative Self-Supervised Learning Method for Radiomic Data}
\author{Zhiyuan Li, Hailong Li, Anca L. Ralescu, \IEEEmembership{Senior Member, IEEE}, Jonathan R. Dillman, \\Nehal A. Parikh, and Lili He 
\thanks{This work was supported by the National Institutes of Health [R01-EB029944, R01-EB030582, R01-NS094200 and R01-NS096037]; Academic and Research Committee (ARC) Awards of Cincinnati Children's Hospital Medical Center. Corresponding author: Lili He (emails: Lili.He@cchmc.org)}
\thanks{Z. Li is with the Imaging Research Center, Department of Radiology, Cincinnati Children’s Hospital Medical Center, and also with the Department of Electronic Engineering and Computer Science, University of Cincinnati, Cincinnati, OH, USA (email: li3z3@mail.uc.edu).}
\thanks{H. Li and L. He are with the Imaging Research Center, Department of Radiology, Artificial Intelligence Imaging Research Center, and Center for Prevention of Neurodevelopmental Disorders, Perinatal Institute, Cincinnati Children’s Hospital Medical Center, and also the Department of Radiology, University of Cincinnati College of Medicine, Cincinnati, OH, USA (emails: hailong.li@cchmc.org; lili.he@cchmc.org).}
\thanks{A. Ralescu is with the Department of Electronic Engineering and Computer Science, University of Cincinnati, Cincinnati, OH, USA (email: ralescal@ucmail.uc.edu).}
\thanks{J. Dillman is with the Imaging Research Center, Department of Radiology, and Artificial Intelligence Imaging Research Center, Cincinnati Children's Hospital Medical Center, and also the Department of Radiology, University of Cincinnati College of Medicine, Cincinnati, OH, USA (email: jonathan.dillman@cchmc.org).}
\thanks{N. Parikh is with the Center for Prevention of Neurodevelopmental Disorders, Perinatal Institute, Cincinnati Children's Hospital Medical Center, and also with the Department of Pediatrics, University of Cincinnati College of Medicine, Cincinnati, OH, USA (email: nehal.parikh@cchmc.org).}
}
\date{}
\maketitle
\begin{abstract}
The computer-aided disease diagnosis from radiomic data is important in many medical applications. However, developing such a technique relies on annotating radiological images, which is a time-consuming, labor-intensive, and expensive process. In this work, we present the first novel collaborative self-supervised learning method to solve the challenge of insufficient labeled radiomic data, whose characteristics are different from text and image data. To achieve this, we present two collaborative pretext tasks that explore the latent pathological or biological relationships between regions of interest and the similarity and dissimilarity information between subjects. Our method collaboratively learns the robust latent feature representations from radiomic data in a self-supervised manner to reduce human annotation efforts, which benefits the disease diagnosis. We compared our proposed method with other state-of-the-art self-supervised learning methods on a simulation study and two independent datasets. Extensive experimental results demonstrated that our method outperforms other self-supervised learning methods on both classification and regression tasks. With further refinement, our method shows the potential advantage in automatic disease diagnosis with large-scale unlabeled data available.
\end{abstract}


\section{Introduction}
\IEEEPARstart{R}{adiomics} is a process that extracts and analyzes high-throughput quantitative features from digital radiographic images acquired by modern medical imaging techniques, such as magnetic resonance imaging (MRI), computed tomography (CT), and positron emission tomography (PET) \cite{lambin2012radiomics}. Even though it was derived from the oncology field, radiomics has been applied to various medical image studies  \cite{ggillies2016radiomics}. Extracted features (referred to as radiomic data/features) are often defined as shape-based, first-, second-, and higher-order statistical descriptors of radiological images, such as signal intensity distribution, tissue/organ morphology/shape, volumetry, and inter-voxel patterns and texture \cite{zwanenburg2020image}. These interpretable radiomic data change with alterations in tissue histology and morphology, thereby, being capable of quantifying phenotypic characteristics in radiological images to aid diagnosis, prognosis, and assessment of response to treatment.

\begin{figure*}
    \centering
    \includegraphics[width=0.88\textwidth]{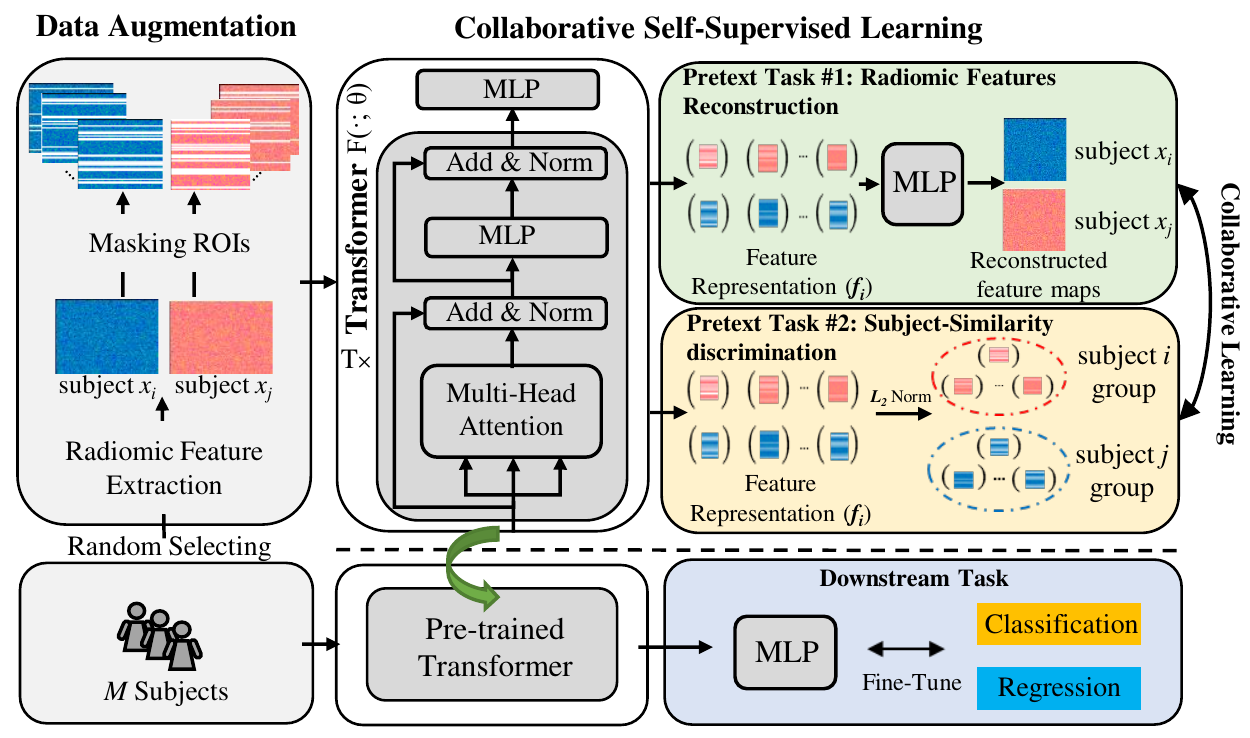}
    \caption{Schematic diagram of the proposed method. For each selected pair of subjects $x_i$ and $x_j$, we randomly masked $k$ out of $N$ ROIs on radiomic feature maps. This process was repeated K times to enlarge the training samples. We then pretrained a Transformer via a joint loss function to solve two pretext tasks i.e., a radiomic feature reconstruction and a subject-similarity discrimination using augmented training samples. Finally, we fine-tuned the pretrained a Transformer to perform downstream tasks. N: Number of Transformer blocks; Add \& Norm: Residual connection and layer normalization; MLP: Multilayer perceptron.}
    \label{framework}
\end{figure*}

The practice of radiomics involves multiple discrete steps, including image acquisition, image segmentation, feature extraction, radiomic data management, and data analysis \cite{lambin2012radiomics,ggillies2016radiomics}. Although each of these complex steps has its own challenges, radiomics continues to rapidly expand as knowledge and tools in each domain have been constantly evolving \cite{ggillies2016radiomics}. Particularly, tools to standardize feature extraction, analytic tools, and open-source platforms (e.g., PyRadiomics \cite{van2017computational} and MIRP \cite{zwanenburg2019assessing}) are available for the research community to automatically extract a large panel of engineered radiomic features from medical images. This largely facilitates the subsequent radiomic data analysis by addressing reproducibility and comparability issues \cite{van2017computational}. Earlier studies applied radiomic data for radiological phenotyping to investigate cancer diagnosis \cite{ou2020radiomics}, survival prediction \cite{isensee2017brain}, malignancy prediction \cite{alahmari2018delta}, recurrence prediction \cite{conti2021radiomics}, and cancer staging \cite{afshar2019handcrafted}. In recent years, radiomic features were also utilized in other medical applications, such as Alzheimer’s disease \cite{feng2020mri,salvatore2021radiomics}, schizophrenia \cite{cui2018disease,park2020differentiating}, and hepatic diseases \cite{park2020radiomics}. Many prior studies have used supervised machine learning models to learn the representations of radiomic features for clinical decision support and to reveal underlying pathophysiology \cite{valdora2018rapid,he2019machine,jeong2019radiomics,beig2020introduction}. Despite the success of those works, some challenges remain in the existing radiomics studies. Supervised learning models typically require a large number of labeled data to achieve reliable and robust performance. However, annotating medical images is a time-consuming, labor-intensive, and expensive process, and it may also involve invasive procedures or long-term follow-up with subjects \cite{li2018novel}, thereby limiting the sample size.  

Recently, self-supervised learning (SSL), a feature representation learning paradigm, was developed to solve the challenges posed by the over-dependence of labeled data. SSL emerged from the fields of natural language processing (NLP) and computer vision \cite{lan2019albert,chen2020simple,he2020momentum}. In the SSL paradigm, we typically pre-train the model by solving pretext tasks in an unsupervised manner, where the data itself provides supervision, and then fine-tune the model by solving real tasks (referred to as downstream tasks) in a supervised manner \cite{kolesnikov2019revisiting}. SSL has been gaining popularity in the medical domains of clinical diagnosis \cite{li2020self}, segmentation \cite{tomar2022self}, and medical text mining \cite{nadif2021unsupervised}. Previous SSL studies have developed a series of pretext tasks with supervisory signals to train the models for learning the latent feature representation, such as the rotation-oriented approach \cite{li2021rotation}, BERT-based model \cite{ji2021does}, Rubik’s cube recovery \cite{zhuang2019self}, anatomical position prediction \cite{bai2019self}, and modality-invariant method \cite{li2020self}. Although these SSL methods achieved excellent performance for various NLP and computer vision tasks, they have not been adapted to radiomic features, whose characteristics are different from text and image data. Applying existing SSL approaches to radiomic data may lead to sub-optimal model performance.

In this study, our objective was to develop a novel SSL technique explicitly for radiomic data. As noted, radiomic data has its own unique characteristics. In practice, in radiomics studies, one would automatically or manually delineate one or more regions of interest (ROIs) on radiological images and then extract high-throughput (e.g., often one-hundreds or more) quantitative features from each ROI, resulting in a radiomic feature map for each subject \cite{zwanenburg2020image}. Such radiomic data (i.e., feature maps) have neither a strong spatial relationship as image data nor a sequential relationship as in NLP language data. The spatial relationship specifies how an object is located in space in relation to other objects in an image, while the sequential relationship specifies how a word is dependent on those that come before or after it in language data. Different from image/language data, ROIs, and quantitative features in radiomic feature maps do not need to follow any specific order, and their positions are invariant to the radiological phenotypes (e.g., tumors, lesions, medical conditions, etc.). In contrast, there exist implicit biological or pathological relationships among different ROIs in radiomic data, which reflect how a ROI is related to other ROIs or pathology  \cite{feng2019self,peng2017multilevel,liu2009liver}. For example, in a long-term follow-up brain development study, volumes (i.e., one of the most common radiomic features) of regional brain ROIs have been observed to follow typical and atypical growth patterns in children \cite{thompson2020tracking}. In another study, simultaneously reduced volume in five ROIs of the gyrus rectus, medial frontal cortex, superior frontal gyrus, inferior frontal gyrus, and subcallosal area were observed and associated with neuropsychiatric symptom scores in a cohort with various neurodegenerative diseases \cite{cajanus2019association}. Using the abdominal MRI image as another example, volume ratio of liver and spleen gradually decreased with worsening hepatic fibrosis \cite{liu2009liver}. 

In this work, by taking advantage of latent biological or pathological relationships in radiomic data, we formulated a pretext task, \textbf{\emph{Radiomic Features Reconstruction}}, by randomly masking or hiding radiomic features from ROIs, and then reconstructing the masked/hidden radiomic features using the visible ones. (\textbf{Figure \ref{framework}}) Instead of grid masks for image data, we explicitly designed row-wise masks by considering the characteristic of radiomic data to hide radiomic features from randomly selected ROIs. In this way, learning to reconstruct the masked radiomic features helps the model understand latent relations among ROIs. The learned radiomic feature representation using such a pretext SSL task may benefit the downstream tasks (e.g., disease classification or outcome regression), in which sufficient annotated data are lacking. 

Nevertheless, these feature representations learned by the \textbf{\emph{Radiomic Features Reconstruction}} pretext task may be sensitive to the position of the masks, i.e., masking different ROIs may produce different mask-dependent feature representations from the same radiomic data. Because our downstream tasks are independent of any pretext task, we formulated another pretext task, \textbf{\emph{Subject-Similarity Discrimination}}, to force the model to learn mask-invariant feature representations. The Subject-Similarity Discrimination task aims to discover the similarity and dissimilarity information from the masked radiomic feature maps and to cluster those from the same subject into one group. In this way, the Subject-Similarity Discrimination task collaboratively aids the Radiomic Features Reconstruction task to learn robust latent feature representations from radiomic data in a self-supervised manner. 

As such, we proposed a novel collaborative SSL approach to solve the challenge of insufficient labeled data for radiomic data by developing a \textbf{\emph{Radiomic Features Reconstruction}} task and a \textbf{\emph{Subject-Similarity Discrimination}} task to collaboratively learn the representative radiomic features from the data itself without any human labeling. A collaborative learning objective was presented for feature representation learning by combining reconstruction and contrastive learning loss functions to collaboratively learn the radiomic features. We further analyzed the proposed objective function using the Bregman divergence in a statistical divergence view. Our proposed collaborative SSL method was specifically designed for radiomic features by considering its unique data characteristic. Then, the SSL-learned feature representation can be used in downstream tasks to fine-tune task-specific supervised learning models. To evaluate the effectiveness of our method, we first propose a simulation study to theoretically compare our methods with other state-of-the-art SSL approaches. We then employed two independent real datasets to test our model for early prediction of abnormal neurodevelopmental outcomes in very preterm infants (VPIs). Our contributions in this paper are outlined as follows: 
\begin{enumerate}
    \item We proposed a novel collaborative SSL approach for radiomic features to address the challenge of insufficient labeled data. By explicitly considering the unique data characteristic of radiomic features, we designed two collaborative pretext tasks, \textbf{\emph{Radiomic Features Reconstruction}} and \textbf{\emph{Subject-Similarity Discrimination}}. Our work is the first to apply SSL techniques to radiomic features.  
    \item With the proposed approach, an integrated learning objective was presented by combining reconstruction and contrastive learning loss functions to collaboratively learn radiomic features. We also provide mathematical derivation to show the properties and advantages of our objective function from a statistical divergence view.
    \item A simulation study was first conducted to show the effectiveness of our method. Substantial experiments on two independent real datasets further demonstrated the superior performance of our proposed model to other state-of-the-art SSL benchmark models for downstream classification/regression tasks. The code of our proposed method is publicly available at \href{https://github.com/leonzyzy/Collaborative_SSL_Radiomic}{https://github.com/leonzyzy/collaborative}.
\end{enumerate}

\section{Related Works}
In this section, we first review the previous works on medical image-based diagnosis from radiomic features and then discuss some related works on the most recent SSL methods.

\subsection{Automatic Disease Diagnosis from Radiomic Data}
Radiomic features have been applied to many medical applications, including neurological disorders and tumors, Alzheimer’s disease \cite{feng2020mri,salvatore2021radiomics}, schizophrenia \cite{cui2018disease,bang2021interpretable}, breast cancer \cite{ou2020radiomics}, and liver disease \cite{starmans2018classification,wei2020radiomics}. Those works \cite{ yue2020machine,he2019machine,peng2018distinguishing} typically involved three main steps: 1) delineate relevant ROIs in MRI images, 2) extract quantitative radiomic features for each ROI, and 3) develop a predictive model with the selected radiomic features for disease diagnosis. For example, Yue et al \cite{yue2020machine} used logistic regression and random forest to predict hospital stay in patients with SARS Cov-2 using CT radiomic features. He et al \cite{he2019machine} stratified the severity of liver stiffness for children and adolescent patients using radiomic features from T2-weighted MRI liver images by developing support vector machine (SVM) models. Brunese et al \cite{brunese2020ensemble} developed an ensemble supervised learning model to detect brain cancer using radiomic features. Peng et al \cite{peng2018distinguishing} combined the SVM model and Isomap (IsoSVM) on brain radiomic features to predict treatment effects after stereotactic radiosurgery. Most prior studies only focused on one or a few ROIs based on prior knowledge to simplify the training procedure of supervised learning models. This was mainly due to the limited sample size of the dataset. Otherwise, supervised learning models can be easily overfitted. In this work, our SSL-based strategy is able to generalize the radiomics to investigate the arbitrary number of ROIs on medical images without substantial prior knowledge. This is achieved by largely improving the issue of supervised learning models’ over-dependence on labeled data.    

\subsection{Self-Supervised Learning}
Recently, SSL has been widely recognized in computer vision and pattern recognition, due to its ability to handle a massive amount of unlabeled data. Various types of self-supervised methods have been developed by designing different pretext tasks with supervisory signals to learn the latent feature representation directly from the data itself. In this section, we discuss some state-of-art SSL models in the domain of computer vision. SSL methods for images can be summarized into three categories, including predictive, reconstructive, and contrastive learning methods. Predictive methods usually apply models to learn the latent features by predicting the pseudo labels, such as rotation prediction \cite{gidaris2018unsupervised},  puzzles solving \cite{noroozi2016unsupervised}, anatomical position prediction \cite{bai2019self}, and 3D distance prediction \cite{spitzer2018improving}. Reconstructive methods aim to learn an encoder mapping input images into latent features and a corresponding decoder to reconstruct the input images from the latent features \cite{liu2021self}. These reconstructive pretext tasks include image denoising with auto-encoders \cite{goodfellow2016deep,wu2020self}, context restoration \cite{chen2019self}, and Rubik's cube \cite{zhuang2019self}. The Contrastive learning method \cite{chen2020simple,he2020momentum} propose tasks to discriminate the subject/instance. The main idea of contrastive learning methods is to train an encoder that embeds the input images into latent representations, and then clusters the representation of the different views from the same images and spread the representation of the views from different images based on a distance estimation (e.g., mutual information). Some representative works are MoCo v1 \cite{he2020momentum}, Invariant \cite{ye2019unsupervised}, SimCLR \cite{chen2020simple}, and BYOL \cite{grill2020bootstrap}. As noted earlier, radiomic features have different data characteristics from images. Although some of these SSL methods for images can be applied to radiomic feature maps, it is very difficult to achieve superior model performance. This was illustrated in our experiments, where we compared the proposed approach with multiple state-of-the-art SSL approaches in this section.

\section{Methodology}
\subsection{Overview}
An overview of the framework is shown in \textbf{Figure \ref{framework}}. Assume that we have a total of $M$ subjects. For each set of MRI images, we can apply MRI preprocessing tools to parcellate the whole MRI images into $N$ different ROIs. Next, we extracted the radiomic features from each ROI using the radiomics pipelines (e.g., PyRadiomics \cite{van2017computational}), thus resulting in a 2D feature map for each subject. Our collaborative SSL approach is an iterative learning procedure. In each iteration, we randomly selected a pair of subjects $x_{i}$ and $x_{j}$ from the training dataset i.e., $\bm{S}=\{x_i\}_{i=1}^{M}$, and masked k random ROIs on individual radiomic feature maps. We repeated this augmentation $K$ times to generate many pretext pseudo-training samples $\Tilde{\bm{x}}_i$. After that, we adapted a Transformer \cite{vaswani2017attention} as a network encoder $F(\cdot; \theta)$ to map the input $\Tilde{\bm{x}}_i$ to a set of latent feature representation $\Tilde{\textbf{f}}_i=\{\Tilde{\text{f}}_{i1},\Tilde{\text{f}}_{i2},\dots,\Tilde{\text{f}}_{iK}\}$. These feature representations are collaboratively optimized via a joint loss function by a radiomic features reconstruction task and a subject-similarity discrimination task. Then, we connected multilayer perceptron (MLP) and output layers to the pretrained Transformer as the model for downstream tasks. Finally, we fine-tuned the whole model on our real downstream task, i.e., cognitive deficits prediction, using the original radiomic features as input to solve the real downstream classification or regression tasks in a supervised fashion using the subjects with the labels. Below, we will elaborate on the radiomic features reconstruction and subject similarity discrimination.

\subsection{Network Encoder}
We applied a Transformer without position encoding as a network encoder in our framework to learn two pretext tasks for embedding the radiomic features. We selected the transformer as the encoder according to the data characteristic of radiomic data. Particularly, radiomic data (i.e., feature maps) are statistical descriptions of individual regions of interest (ROIs) on the images. Though the radiomic data appear to be 2D maps, they have no strong spatial relationship on the feature maps. Thus, different from most of the other existing SSL methods, convolutional neural networks (CNNs)-based models are inappropriate to work as feature encoder for radiomic data. Actually, radiomic data are a set of vectors. In this sense, radiomic feature maps are similar to natural language data, where each word is commonly represented by a vector (a.k.a., word embeddings) and the whole sentence becomes a set of vectors. There exist implicit pathological or biological relationships among the different brain regions (i.e., ROIs) in the radiomic feature map, which is presented by how one a ROI is related to all other ROIs \cite{peng2017multilevel,liu2009liver}.
As such, Transformer \cite{vaswani2017attention}, more precisely, the built-in self-attention mechanism, can efficiently capture how individual ROI is related to all other ROIs. Thus, we believe that Transformer as the encoder network can learn better feature embeddings compared to other encoders (e.g., CNNs). Furthermore, radiomic data does not have a strong sequential relationship as in language data. That is, the order of brain regions (i.e., ROIs positions) is invariant to the radiological phenotypes (e.g., tumors, lesions, medical conditions, etc.). Accordingly, we removed the position encoding module in the typical Transformer to avoid the model to learn/remember the order of the ROIs that are invariant to downstream tasks.

\subsection{Radiomic Features Reconstruction Task}
To discover the latent relation among radiomic features from different ROIs, we designed the radiomic features reconstruction task. The reconstruction pretext task is similar to the masked autoencoder (MAE) \cite{he2022masked}, which masks out a large random subset of image patches and pretrains a network to reconstruct the masked patches. The major difference is that, for the MRI radiomic data, we designed to mask several ROIs from the input radiomic feature maps and conduct a radiomic feature reconstruction task to help the model understand the latent relations among ROIs. To reach that, the input $\Tilde{\bm{x}}_i$ is first fed to a Transformer to produce the high-level features $\Tilde{\textbf{f}}_i$, and then $\Tilde{\textbf{f}}_i$ is fed into a multi-layer perceptron (MLP) to produce the reconstructed radiomic feature map $\hat{\bm{x}}_i$ with the same feature dimension size as $\Tilde{\bm{x}}_i$. We defined our reconstruction loss for the first pretext task as a convex linear combination of distance metrics of square error and JS-Divergence. The square error measures the Euclidean distance between reconstructed radiomic feature map $\hat{\bm{x}}_i$ and ground-truth radiomic feature map $\Tilde{\bm{x}}_i$. However, the square error only considers the distance metrics within the Euclidean space, while the probabilistic distance measure between the probability density function of $\hat{\bm{x}}_i$ and $\Tilde{\bm{x}}_i$ the was ignored. This may lead to an inaccurate distance measure for radiomic features that are high dimension \cite{saito1994new, wang2005euclidean, bai2013graph}. To better learn the probability distribution between $\hat{\bm{x}}_i$ and $\Tilde{\bm{x}}_i$i, we designed a convex linear combination of square error and JS-Divergence using a weighting factor $\beta\in [0,1]$ to control each loss importance, such that
\begin{align}
    \mathcal{L}_{r} &=\sum_{i\in M}\sum_{j\in K}\beta \|x_i-\hat{x}_{ij(k,N)}\|^2 \nonumber \\
    &+ (1-\beta)d_{\text{JS}}\left(p(x_i)\|p(\hat{x}_{ij(k,N)})\right)
    \label{eq1}
\end{align}
where $\hat{x}_{ij(k,N)}\in \hat{\bm{x}}_i$ represents the $j_{\text{th}}$, $j\in K$, reconstructed radiomic feature map of $\hat{\bm{x}}_i$. $d_{\text{JS}}$ indicates the Jensen Shannon (JS) divergence \cite{lin1991divergence} between two probability distributions of $x_i$ and $z$, which is defined as:
\begin{align}
    d_{\text{JS}}(p(x_i)\|p(\hat{x}_{ij(k,N)})) &= \frac{1}{2}d_{\text{KL}}\left(p(x_i)\|p(z)\right) \nonumber \\
    &+ \frac{1}{2}d_{\text{KL}}\left(p(\hat{x}_{ij(k,N)})\|p(z)\right)
    \label{eq2}
\end{align}
where $p(z) = \frac{1}{2}\left[p(x_i) + p(\hat{x}_{ij(k,N)})\right]$ and $d_{\text{KL}}$ indicates the  Kullback–Leibler (KL) divergence \cite{kullback1951information}. In practice, we used a SoftMax to generate the probabilities. $\beta \in [0,1]$ of \textbf{Eq (\ref{eq1})} denotes a weighting factor to leverage the square error loss and JS-Divergence as a convex linear combination. We also analyzed effects of $\beta$ in our ablation study section.

\subsection{Subject-Similarity Discrimination Task}
To learn the subject-similarity information, we designed the subject-similarity discrimination task to learn the mask-invariant features by clustering the similar augmented samples and separating the dissimilar ones. As shown in \textbf{Figure} \ref{framework}, $\Tilde{\textbf{f}}_i=\{\Tilde{\text{f}}_{i1},\Tilde{\text{f}}_{i2},\dots,\Tilde{\text{f}}_{iK}\}$ denotes a set of embedding of masked maps, which is obtained by the Transformer $F(\cdot; \theta)$ for $\Tilde{\bm{x}}_i$. We first apply $L_2$ norm to normalize each $\Tilde{\text{f}}_{ij} \in \Tilde{\textbf{f}}_i$, i.e., $\|\Tilde{\text{f}}_{ij}\|_2=1$. Then, the negative log-likelihood for a similar pair $(\Tilde{\text{f}}_{iv},\Tilde{\text{f}}_{ij}), i\ne j$, is defined as follows:
\begin{align}
    \ell_{d}(\Tilde{\text{f}}_{iv},\Tilde{\text{f}}_{ij})&=-\log \frac{\sum_{v,j\in K}\exp(\Tilde{\text{f}}_{iv}^T\Tilde{\text{f}}_{ij}/\tau)}{\sum_{m\in M}\sum_{j\in K}\exp(\Tilde{\text{f}}_{iv}^T\Tilde{\text{f}}_{mj}/\tau)}
    \label{eq3}
\end{align}
where $\tau$ is the temperature scale parameter, which is set to 0.1 empirically \cite{chen2020simple}. $\Tilde{\text{f}}_{iv}^T\Tilde{\text{f}}_{ij}$ denotes the inner product (e.g., Cosine similarity) between $\Tilde{\text{f}}_{iv}$ and $\Tilde{\text{f}}_{ij}$. Note, $\ell_{d}(\Tilde{\text{f}}_{iv},\Tilde{\text{f}}_{ij})$ is asymmetric, i.e., $\ell_{d}(\Tilde{\text{f}}_{iv},\Tilde{\text{f}}_{ij})\ne \ell_{d}(\Tilde{\text{f}}_{ij},\Tilde{\text{f}}_{iv})$. Thus, the overall discrimination loss is described as:
\begin{align}
    \mathcal{L}_{d} = \frac{1}{2M}\sum_{i\in M}[\ell_{d}(\Tilde{\text{f}}_{iv},\Tilde{\text{f}}_{ij})+\ell_{d}(\Tilde{\text{f}}_{ij},\Tilde{\text{f}}_{iv})]
    \label{eq4}
\end{align}
where $\Tilde{\text{f}}_{iv}$ stays close to $\Tilde{\text{f}}_{ij}$ in the embedding space when $\mathcal{L}_{d}$ is optimized over time by training the network.

\subsection{Representation Learning Objective }
\subsubsection{Loss Function}
We define the total learning objective function as the weighted linear combination of a radiomic features construction task and a subject-similarity discrimination task. The objective loss function is defined as follows:
\begin{align}
     \mathcal{L}^* &= \mathcal{L}_{r} +\lambda \mathcal{L}_{d}
     \label{eq5}
\end{align}
where $\lambda$ is a weighting factor, which controls the importance of the subject-similarity discrimination task. We empirically set $\lambda$ to 1 \cite{li2021rotation}. We also compared the model performance based on different $\lambda$ in the ablation study. 

\subsubsection{Statistical Divergence View}
We show the equivalent relation and properties of our learning objective loss using the Bregman divergence \cite{bregman1967relaxation}. Let $\psi:\Omega\to \mathbb{R}$ be a strictly convex function, which is continuously differentiable on a closed convex set $\Omega$. Given two vectors of $p$ and $q$, the Bregman divergence between $p$ and $q$ is described as:
\begin{align}
    d_{\psi}(p,q) &= \psi(p) - \psi(q) - \nabla \psi(q)^T(p-q)
    \label{eq6}
\end{align}
where $\psi$ indicates a generating function that is convex and $\nabla \psi(q)$ represents the gradient of $\psi(q)$. In below, we show that the learning objective $\mathcal{L}^{*}$ can be formulated into $d^{*}_{\psi}$ by rewriting $\mathcal{L}_{r}$ and $\mathcal{L}_{d}$ to $d^{r}_{\psi}$ and $d^{d}_{\psi}$, respectively.  

For $\mathcal{L}_{r}$ of \textbf{Eq (\ref{eq1})}, let $\psi(x_i)$ be $\langle x_i , \; x_i \rangle$, which is continuous differentiable and strictly convex in $\mathbb{R}$. Hence, the Bregman divergence $d^{(1)}_{\psi}(x_i,\hat{x}_{ij})$ is defined as follows:
\begin{align}
    d_{\psi}^{(1)}\left(x_i,\hat{x}_{ij}\right) &= \|x_i\|^2-\|\hat{x}_{ij}\|^2
    -\langle x_i -\hat{x}_{ij},\; 2\hat{x}_{ij} \rangle \nonumber \\
    &= \|x_i-\hat{x}_{ij}\|^2
    \label{eq7}
\end{align}
where $\langle \; \rangle$ denotes the inner product. Now, let $\psi(x_i)=\sum_{h\in H}p_{x_{ih}}\log p_{x_{ih}}$ be a strictly convex function, which is continuously differentiable in $\mathbb{R}$. The corresponding Bregman divergence $d^{(1)}_{\psi}(x_i,z)$ is
\begin{align}
    d^{(2)}_{\psi}(x_i,z) &= \sum_{h\in H}p_{x_{ih}}\log \frac{p_{x_{ih}}}{p_{z_{h}}}-\log \exp\left[\sum_{h\in H}(p_{x_{ih}}-p_{z_{h}})\right] \nonumber \\
    & = d_{\text{KL}}\left(p(x_i)\|p(z)\right)
    \label{eq8}
\end{align}
where $\sum_{h\in H}p_{x_{ih}}=\sum_{h\in H}p_{z_{h}}=1$. Hence, based on \textbf{Eq  (\ref{eq7})-(\ref{eq8})}, we finalize $\mathcal{L}_{r}$ as 
\begin{align}
    \mathcal{L}_{r} = \sum_{i\in M}\sum_{j\in K} \beta d_{\psi}^{(1)}(x_i,\hat{x}_{ij}) + (1-\beta) d^{(2)}_{\psi}(p(x_i),p(z))
    \label{eq9}
\end{align}

Next, we show $\mathcal{L}_{d}$ of \textbf{Eq (\ref{eq3})} belongs to the Bregman divergence. We first formulate $\ell_{d}$ into a distance form without loss of generality. 
\begin{align}
    \ell_{d}(\Tilde{\text{f}}_{iv},\Tilde{\text{f}}_{ij})&=-\log \frac{\sum_{m\in M}\sum_{j\in K}\exp(\Tilde{\text{f}}_{iv}^T\Tilde{\text{f}}_{mj}/\tau)}{\sum_{j\in K}\exp(\Tilde{\text{f}}_{iv}^T\Tilde{\text{f}}_{ij}/\tau)} \nonumber \\
    &= \log\left[1+\sum_{m\ne i}\sum_{j}\exp(\Tilde{\text{f}}_{iv}^T\Tilde{\text{f}}_{mj}/\tau-\Tilde{\text{f}}_{iv}^T\Tilde{\text{f}}_{ij}/\tau)\right] \nonumber
\end{align}
Using the Taylor approximation:
\begin{align}
    \ell_{d}(\Tilde{\text{f}}_{iv},\Tilde{\text{f}}_{ij}) &\approx \sum_{m\ne i}\sum_{j}\exp(\Tilde{\text{f}}_{iv}^T\Tilde{\text{f}}_{mj}/\tau-\Tilde{\text{f}}_{iv}^T\Tilde{\text{f}}_{ij}/\tau) \nonumber \\
    &\approx 1+ \sum_{m\ne i}\sum_{j}\Tilde{\text{f}}_{iv}^T\Tilde{\text{f}}_{mj}/\tau-\Tilde{\text{f}}_{iv}^T\Tilde{\text{f}}_{ij}/\tau \nonumber \\
    &\propto \sum_{m\ne i}\sum_{j} d_{\psi}^{(1)}(\Tilde{\text{f}}_{iv},\Tilde{\text{f}}_{mj})-d_{\psi}^{(1)}(\Tilde{\text{f}}_{iv},\Tilde{\text{f}}_{ij})
    \label{eq10}
\end{align}
Hence, based on \textbf{Eq (\ref{eq10})}, $\mathcal{L}_{d}$ is finalized as 
\begin{align}
    \mathcal{L}_{d} &\propto d^{*}_{\psi}=\frac{1}{2M}\sum_{i}\sum_{m\ne i}\{ \sum_{j}d_{\psi}^{(1)}(\Tilde{\text{f}}_{iv},\Tilde{\text{f}}_{mj})-d_{\psi}^{(1)}(\Tilde{\text{f}}_{iv},\Tilde{\text{f}}_{ij}) \nonumber \\
    &+\sum_{v}d_{\psi}^{(1)}(\Tilde{\text{f}}_{ij},\Tilde{\text{f}}_{mv})-d_{\psi}^{(1)}(\Tilde{\text{f}}_{ij},\Tilde{\text{f}}_{iv})\}
    \label{eq11}
\end{align}

Based on \textbf{Eq (\ref{eq6}-\ref{eq11})}, the minimizing the learning objective loss $\mathcal{L}^*$ is equivalent to minimize two Bregman divergences. Thus, $\mathcal{L}^*$ also follows the mathematical properties of non-negativity/positivity, convexity, linearity, duality, generalized Pythagorean theorem, and others of the Bregman divergence. We also compare our $\mathcal{L}^*$ with other difference losses in the ablation study.

\section{Data and Experiments}
\subsection{Datasets}
\subsubsection{Simulated Dataset}
We simulated the radiomic features datasets based on Corso et al \cite{corso2021challenge} to theoretically investigate the different SSL methods with different training sample sizes and the difficulty level of the class separation. Our simulation is based on the statistics information of the MRI images. First, we obtained radiomic feature maps from the MRI brain images, and calculated correlation matrix, skewness, and kurtosis, which capture important information about the real radiomic features. Next, we simulated non-Gaussian multivariate distributions using the computed statistics (e.g., correlation, skewness, kurtosis). In addition, considering that the range of simulated radiomic features may scale a lot, therefore, we re-scaled each simulated feature to the original features ranges of the CINEPS dataset (\textbf{see below}) using the same re-scaling method as \cite{corso2021challenge}. We further randomly assign label ‘1s’ to the 50\% of data and ‘0s’ to the rest of the data.

The separation noise $\phi$ for the $i_{\text{th}}$ feature is defined as follows:
\begin{align}
    \phi^{'0'}(f_{i}^*)&=f_{i}^*-\frac{\bar{f}_{i}^*}{\theta_i}, i\in[0,\infty] \nonumber \\
    \phi^{'1'}(f_{i}^*)&=f_{i}^*+\frac{\bar{f}_{i}^*}{\theta_i}, i\in[0,\infty] \nonumber 
\end{align}
where $\phi^{'0'}$. $\phi^{'0'}$ are the noises for label $'0'$ and $'1'$, respectively. $\bar{f}_{i}^*$ denotes the mean of re-scaled feature $f_{i}^*$ and $\theta_i$ controls the difficulty of the class separation. For each selected sample size, we generated two different separation noises for the selected ROIs associated with the labels ($'0'$ and $'1'$) to control the difficulty level of classification tasks. Afterward, we added the Gaussian noise from a standard normal distribution to the simulated dataset.

\subsubsection{CINEPS Dataset}
Cincinnati Infant Neurodevelopment Early Prediction Study (CINEPS) \cite{parikh2021perinatal} collected T2-weighted MRI brain images from 362 very preterm infants ($\le$32 weeks gestational age) at Cincinnati Children’s Hospital Medical Center (CCHMC). All infants with congenital or chromosomal anomalies that impact the central nervous system were excluded. Each subject was imaged at 39-44 weeks postmenstrual age on the same 3T Philips Ingenia scanner using a 32-channel head coil at CCHMC. Acquisition parameters for axial T2-weighted turbo spin-echo sequence is set as repetition time (TR)=8300 ms, echo time (TE)=166 ms, FA=90°, resolution 1.0 × 1.0 × 1.0 mm3, and time 3:53 min. For each subject, a Bayley Scale of Infant and Toddler Development, Third Ed. (Bayley III) cognitive scores \cite{bayley2006bayley} that reflect neurodevelopment were assessed at 2 years corrected age. We tested our models using a classification task of distinguishing subjects at high- (test score$\le$85) or low-risk (test score$>$85), and a regression task of predicting Bayley III scores (continuous). 

\begin{figure}[ht]
    \centering
    \includegraphics[width=8.00cm]{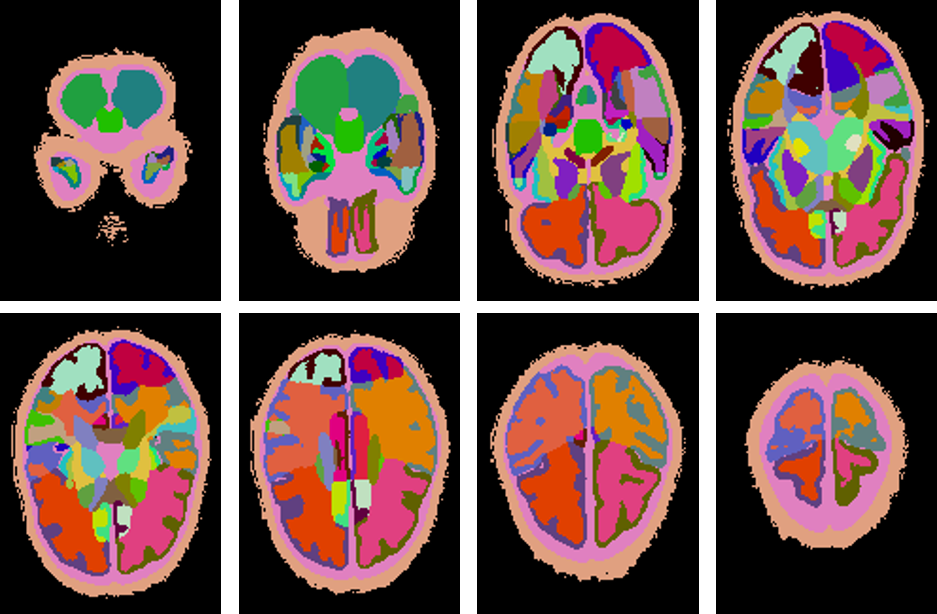}
    \caption{Axial cross-section of the neonatal volumetric brain atlas (40 gestational weeks) used in the dHCP pipeline from inferior to superior slices. For each subject, the pipeline segmented the whole brain into 87 regions of interest (ROIs).}
    \label{dHCP}
\end{figure}

We applied dHCP (Developing Human Connectome Project) pipeline to segment the whole brain image into 87 region-of-interests (ROIs) based on an age-matched neonatal volumetric atlas \cite{gousias2012magnetic, makropoulos2018developing}. Briefly, the pipeline first segmented the T2 MRI image data into 9 tissue classes (e.g., cortical grey matter, white matter, ventricle) using the Draw-EM (Developing brain Region Annotation with Expectation-Maximization) algorithm \cite{makropoulos2018developing}, and then registered the labeled neonatal atlases with 87 ROIs to the subject using a multi-channel registration approach. The neonatal atlas was created by manually labeling T1 and T2 brain MRI images from 20 neonatal subjects \cite{gousias2012magnetic, makropoulos2018developing}. The full list of 87 ROIs can be found in the original paper. \textbf{Figure \ref{dHCP}} illustrates the brain structure of ROIs of the brain atlas. 

After segmenting each T2-weighted brain image into 87 ROIs using the dHCP pipeline, we extracted 100 radiomics features from each ROI using the PyRadiomics pipeline \cite{van2017computational}, therefore resulting in a 2D radiomic feature map for each subject. Specifically, we utilized PyRadiomics [version: 3.0.1] to extract 100 features from feature classes shape [10 features], first order [19 features], Gray Level Co-occurrence Matrix (GLCM) [24 features], Gray Level Size Zone Matrix (GLSZM) [16 features], Gray Level Run Length Matrix (GLRLM) [16 features], Neighbouring Gray Tone Difference Matrix (NGTDM) and Gray Level Dependence Matrix (GLDM) [14 features]. We first conduct image intensity Z-transform normalization and isotropic image voxel resampling (i.e., 1.0 x 1.0 x 1.0 $\text{mm}^3$) on the original unfiltered images using ‘sitkBSpline’ as the interpolator. This was followed by quantitative metrics calculation using the width of the histogram bin as 25.

\subsubsection{COEPS Dataset}
COEPS Dataset: Columbus Early Prediction Study (COEPS) dataset includes 69 very preterm infants from Nationwide Children’s Hospital (NCH). All infants with congenital or chromosomal anomalies that impact the central nervous system were excluded. Subjects were scanned at 38-43 weeks PMA on the same 3T MRI scanner (Skyra; Siemens Healthcare) with a 32-channel pediatric head coil at NCH. Acquisition parameters for axial T2-weighted fast spin-echo sequence are TR = 9500 ms, TE = 147 ms, FA = 150°, resolution 0.93 × 0.93 × 1.0 mm3 , and time 4:09 min. Cognitive Bayley III sub-scores were also collected for each subject at 2 years corrected age. We implemented the same radiomic feature processing pipeline to calculate radiomic feature maps as the CINEPS dataset.

\subsection{Experimental Setting}
\subsubsection{Model Implementation Details}
As illustrated in \textbf{Figure \ref{framework}}, we empirically set 3 Transformer blocks in the network encoder $F(\cdot; \theta)$ in the tensor dimension of each radiomic feature map of $\Tilde{\bm{x}}_i$ is $M\times N\times 100$, and the number of heads in the multi-head attention layer is set to 8 as default. Two MLP layers are included in $F(\cdot; \theta)$ where the first MLP layer (100 nodes) with the residual norm connection is to produce a $N\times 100$ attention map $\Tilde{\bm{a}}_i$, and it forwards to the second MLP layer (8 nodes) to produce the embedding of masked maps $\Tilde{\textbf{f}}_i$ with the dimension of $N\times 8$. Then $\Tilde{\textbf{f}}_i$ is connected with a third MLP layer (100 nodes) and a $L_2$ normalization layer to jointly learn the radiomic features reconstruction task and the subject similarity discrimination task, respectively. In each iteration of the proposed approach, we randomly selected a pair of subjects to perform a random-masking $k \in [1,30]$ with K=50 repetitions on the radiomic feature maps. Same as \cite{li2021rotation}, the network is optimized using the Adam optimizer with a learning rate of 0.001 and a weight decay of 0.001. Note, the position encoding of the Transformer is discarded given no spatial or sequential relationships among the radiomic features. Finally, we fine-tune the pretrained $F(\cdot; \theta)$ by adding a MLP layer (100 nodes) with a Softmax to perform the supervised downstream classification task using a weighted cross-entropy loss and regression tasks using the MSE loss function. We trained both pretext and downstream tasks for 500 epochs, and the batch-size is set to 8 for each epoch. The whole framework was implemented using python 3.8, Scikit-Learn 0.24.1, Pytorch 1.9.1, and Cuda 11.1 with a NVIDIA GeForce GTX 1660 SUPER GPU. 

\subsubsection{Competing SSL Approaches}
We compared our method with other self-supervised methods, including predictive based methods (e.g., Rotation Prediction \cite{gidaris2018unsupervised}, Puzzle Solving \cite{noroozi2016unsupervised}), reconstructive based methods (e.g., MLM \cite{lan2019albert}), and contrastive based methods (e.g., Moco v1 \cite{he2020momentum}, Invariant \cite{ye2019unsupervised}, SimCLR \cite{chen2020simple}). For a fully-supervised baseline model, we trained the same Transformer and MLP without pre-training on any pretext tasks. We used additional training epochs to ensure the model has been sufficiently optimized. We perform these methods using the code that has been released in previous publications. For \cite{gidaris2018unsupervised}, we transformed the input into four different degrees, i.e., 0°, 90°, 180°, and 270°, and trained a network for predicting these four rotations. We permutated the input into 32 different patch combinations in \cite{noroozi2016unsupervised}, then trained a network to classify the permutation order index. To compare with reconstructive-based methods  \cite{devlin2018bert}, we performed the same training strategy as the radiomic features reconstruction task, the only difference is the optimization method that we only trained the network with MSE loss to reconstruct the hidden radiomic features. To compare with contrastive methods \cite{he2020momentum,chen2020simple,ye2019unsupervised}, we applied the same data augmentation techniques to perform the unsupervised learning on embedding features space, the only difference is that we changed the encoder CNN backbone to the Transformer of our method. To have a fair comparison, we trained all SSL models on the Transformer with the same network architectures, batch size, and training optimizer, including the learning rate and the number of training epochs in both self-supervised stages (pretext task) and fine-tuning stage (downstream task). Since the fully-supervised baseline model does not contain a pre-training step on  pretext tasks, we opted to train the model with 2000 epochs to ensure the fair comparison with other SSL models.

\subsubsection{Model Evaluation}
We evaluated our method in both regression and classification metrics. We used mean absolute error (MAE) and $R^2$ to evaluate the predictive performance of our Bayley III cognitive score. For risk stratification (i.e., binary classification), we used balanced accuracy (BA), sensitivity (SEN), specificity (SPE), and the area under the receiver operating characteristic (ROC) curve (AUC) to evaluate classification performance. 

We conducted a nested 10-fold cross-validation consisting of an inner loop and outer loop on the CINEPS dataset to assess internal validation. In the outer loop, we first separated the dataset into training data, validation data, and testing data in each of the 10 iterations. We then optimized the model on training-validation data without any information leaking from testing data. This validation process was repeated 100 times to report mean and standard deviation (SD) to ensure model reproducibility. We tested the optimized model trained from the CINEPS dataset using an unseen independent COEPS dataset.  

To show the statistical significance of the model comparison, we conducted the non-parametric Wilcoxon test for all statistical inference testing based on the $\alpha=0.05$ level. All statistical tests were conducted in R-4.0.3 (RStudio, Boston, MA, USA).

\subsection{Model Comparison Results with Simulation Data}
\begin{figure}[ht]
    \centering
    \includegraphics[width=8.85cm]{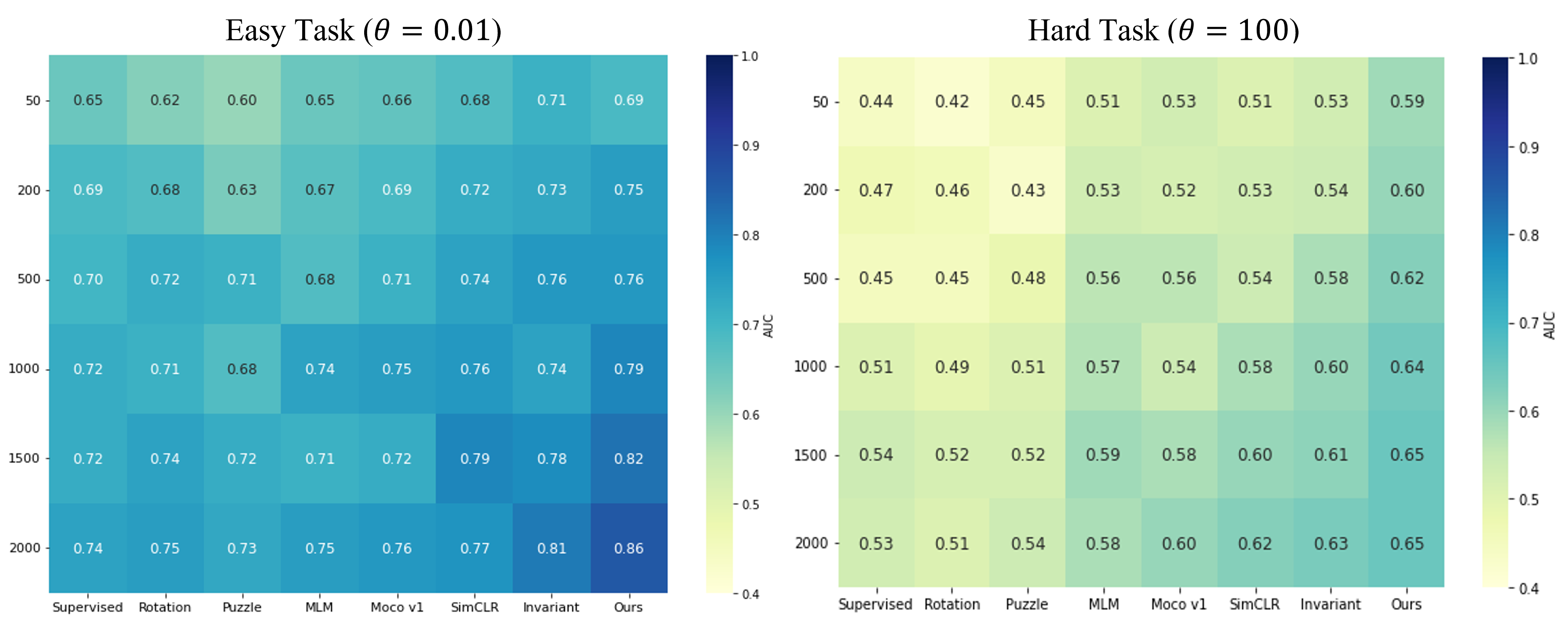}
    \caption{Classification performance comparison on the simulated radiomic dataset with different training sample sizes N and feature noises $\theta$ (high/low). Our collaborative self-supervised learning (SSL) method showed the best classification results as N increases in both two tasks. The data in the table are area under the receiver operating characteristic curve (AUC) values.}
    \label{sim}
\end{figure}

We compared our method with other SSL methods using the simulated datasets with different training sample sizes N and difficulty levels of the classification task. We generated varying numbers of synthetic data samples (N=50, 200, 500, 1000, 1500, 2000) with two difficulty levels, “easy task” ($\theta$ = 0.01) and “hard task” ($\theta$ =100), respectively. The results are shown in $\textbf{Figure \ref{sim}}$. When N=50, each SSL method achieved an AUC above 0.60 on the dataset for the easy task but a lower AUC for the hard task. As N increases, the performance of each method increases since more training samples can improve model performance. Notably, our method can achieve the best classification performance with the highest AUC on both datasets in two tasks. The Invariant approach \cite{ye2019unsupervised} obtained the second best in both tasks. Rotation \cite{gidaris2018unsupervised} and jigsaw puzzle \cite{noroozi2016unsupervised} had an inferior performance on the dataset in both tasks. The simulation results theoretically showed the effectiveness of the proposed collaborative SSL method. 

\subsection{Model Comparison Results with CINEPS Dataset}
\begin{figure}[ht]
    \centering
    \includegraphics[width=8.85cm]{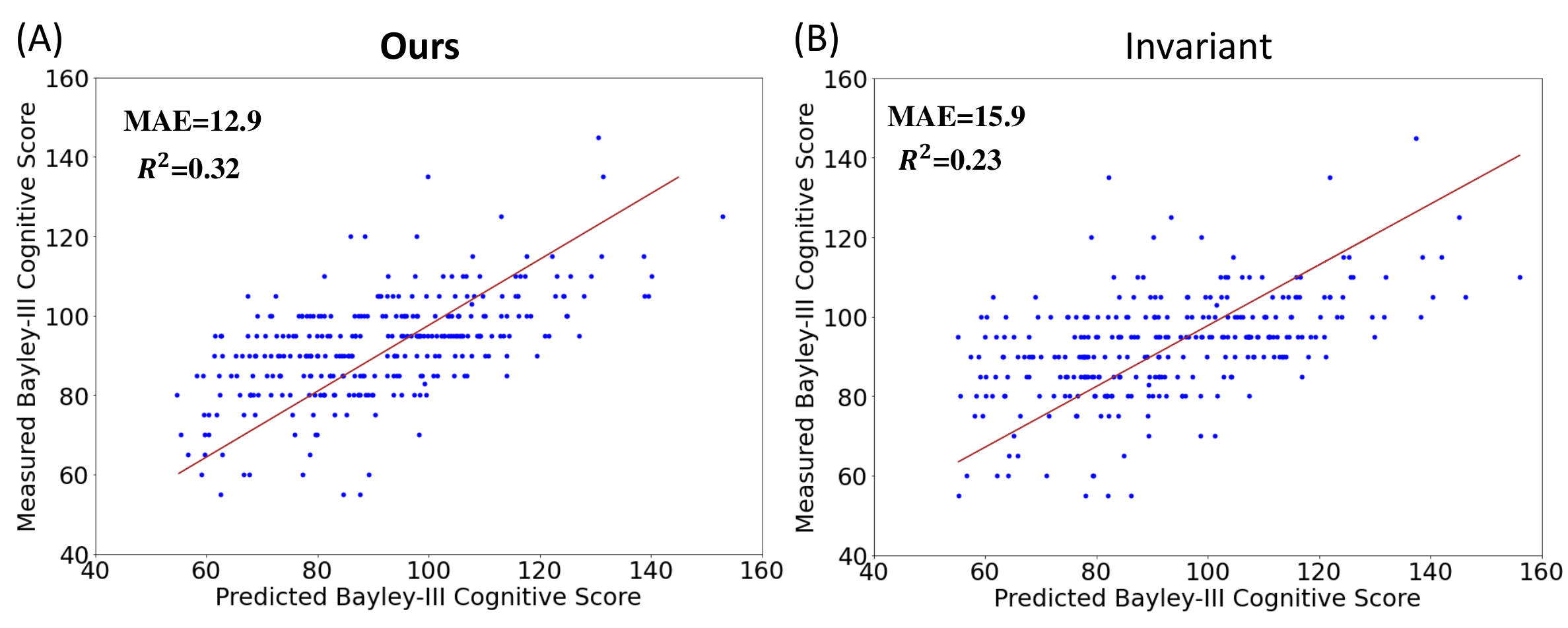}
    \caption{Bayley III test score regression for early prediction of cognitive deficits in VPIs at 2 years of corrected age on the CINEPS dataset. We included the top 2 self-supervised methods based on the AUC of cognitive deficits risk stratification. We report the mean absolute error (MAE) and $R^2$ to evaluate the regression performance.}
    \label{reg1}
\end{figure}

We evaluate the performance for cognitive deficits risk classification and Bayley III Cognitive score regression. As shown in \textbf{Table \ref{Tab1}}, our methods achieved the best classification performance among other competing SSL methods. Compared to the second-best method Invariant \cite{ye2019unsupervised}, our method significantly improved classification performance by around 5.7\% (p$<$0.001) on balanced accuracy, 9.4\% (p$<$0.001) on sensitivity, 1.9\% (p$<$0.001) on specificity, and 0.07 (p$<$0.001) for AUC. The predictive-based methods, \cite{gidaris2018unsupervised} and \cite{noroozi2016unsupervised} had comparable results as the supervised learning baseline. Our method outperformed the supervised learning baseline, 76.3\% vs 66.8\% on balanced accuracy and 0.78 vs 0.66 for AUC. For Cognitive score prediction (\textbf{Figure \ref{reg1}}), our method achieved the best regression performance with a MAE of 12.9 and $R^2$ of 0.32 (p$<$0.001), showing a significant correlation between the predicted cognitive score and the actual cognitive assessment. Compared with the second-best model Invariant \cite{ye2019unsupervised}, our method significantly achieves a lower MAE (p$<$0.001) and a higher $R^2$ (p$<$0.001). These results further demonstrated the effectiveness of our method.
\begin{table}[ht]
    \centering
    \caption{Model comparison on the CINEPS dataset for early risk stratification of very preterm infants at high risk for cognitive deficits. Network encoder: Transformer. Experimental results are represented as mean (SD).}
    \begin{tabular}{ccccc}
     \hline
     	& BA (\%) & SEN (\%) & SPE (\%) & AUC\\
     \hline
     Supervised \cite{vaswani2017attention} & 67.4(4.7) & 66.2(6.7) & 68.5(6.2) & 0.67(0.07) \\
     \hline
     Moco v1 \cite{he2020momentum} &70.3(4.8) & 68.4(7.6) & 72.2(6.4) & 0.69(0.06) \\
     MLM \cite{devlin2018bert} &68.9(5.2) & 66.7(7.5) & 71.2(5.9) & 0.68(0.07) \\
     SimCLR \cite{chen2020simple} &69.3(5.2) & 67.5(7.5) & 71.1(6.8) & 0.68(0.07) \\
     Invariant \cite{ye2019unsupervised} &70.6(4.9) & 66.4(7.2) & 74.8(5.8) & 0.71(0.07) \\
     Jigsaw Puzzle \cite{noroozi2016unsupervised} &64.8(4.4) & 63.1(5.4) & 66.5(5.5) & 0.62(0.06) \\
     Rotation \cite{gidaris2018unsupervised} &66.2(4.1) & 64.2(5.3) & 68.1(5.2) & 0.64(0.06) \\
     \textbf{Ours}  & \textbf{76.3(4.9)} & \textbf{75.8(6.9)} & \textbf{76.7(6.1)} & \textbf{0.78(0.07)} \\
     \hline
    \end{tabular}
    \label{Tab1}
\end{table}

\subsection{Model Comparison Results With COEPS Dataset}
To show the generalization of our method, we performed an external validation on the COEPS dataset by training all the models on the CINEPS dataset. The results are shown in \textbf{Table \ref{Tab4}}. Moco v1 \cite{he2020momentum} and SimCLR \cite{chen2020simple} had the lowest performance in classifying the high-risk group of cognitive deficits, and predicting the Bayley III cognitive score. It is observed that Invariant \cite{ye2019unsupervised} showed very promising predictive capability with the highest sensitivity and precisely predicted Bayley III cognitive score. Again, our method can achieve the best performance in most of the tasks. It had a higher risk classification performance (1-2\% improvement in AUC) than Invariant \cite{ye2019unsupervised}. For regression, our method also demonstrated a strong predictive capability. Significant linear relationships were observed between the predicted Bayley III cognitive scores and the actual Bayley III scores $R^2=0.30$ (p$<$0.001) for cognitive development. These external results further illustrated the generalization capability of our method.
\begin{table}[ht]
    \centering
    \caption{Model comparison self-supervised learning methods on the COEPS dataset for early prediction of very preterm infants at high-risk for cognitive deficits. All models are optimized on the CINEPS dataset. BA, SEN, and SPE are represented with UNIT: \%.}
    \begin{tabular}{ccccccccc}
     \hline
      & BA  & SEN & SPE  & AUC & MAE & $R^2$\\
     \hline
      Supervised \cite{vaswani2017attention} & 62.2 & 60.0 & 64.4 & 0.63 & 23.4 & 0.14 \\
      \hline
      Moco v1 \cite{he2020momentum} & 64.0 & 50.0 & \textbf{78.0} & 0.62 & 25.3 & 0.12\\
      SimCLR \cite{chen2020simple} & 63.0 & 60.0 & 66.1 & 0.65 & 22.1 & 0.17 \\
      Invariant \cite{ye2019unsupervised} & 68.9 & \textbf{70.0} & 67.8 & 0.69 & 18.5 & 0.25 \\
      MLM \cite{devlin2018bert}  & 64.7 & 60.0 & 69.5 & 0.65 & 21.3 & 0.14\\
      \textbf{Ours} & \textbf{71.4} & \textbf{70.0} & 72.9 & \textbf{0.70} & \textbf{16.5} & \textbf{0.30}\\
     \hline
    \end{tabular}
    \label{Tab4}
\end{table}

\subsection{Ablation Study of The Proposed SSL Model}
\subsubsection{Comparison of Each Individual Collaborative Pretext Task}
Our method formulated two Collaborative Pretext tasks to learn radiomic data. To investigate individual pretext tasks that work collaboratively, we analyzed the effects of each pretext task for identifying cognitive deficits using the CINEPS dataset. As shown in \textbf{Table \ref{Tab6}}, model trained with the subject-similarity discrimination task alone achieved better classification performance than with the radiomic features reconstruction task alone, i.e., 72.0\% vs 69.8\% for BA, 0.72 vs 0.70 for AUC. Such phenomenon has been observed in \cite{grill2020bootstrap,li2021rotation}, indicating the contrastive-based pretext may perform better than other hand-crafted predictive-based pretext tasks. Notably, our method, which collaboratively used two pretext tasks, can achieve a higher classification performance than the two individual tasks with 0.08 (p$<$0.001) and 0.06 p$<$0.001) improvement on AUC, respectively.
\begin{table}[ht]
    \centering
    \caption{Comparison of Each Individual Collaborative Task for the cognitive deficits risk stratification on the CINEPS dataset.}
    \begin{tabular}{ccccc}
     \hline
      & BA (\%) & SEN (\%) & SPE (\%) & AUC\\
      \hline
      Reconstruction & 69.8(5.1) & 67.2(7.5) & 72.3(6.2) & 0.70(0.07) \\
      Discrimination & 72.0(5.5) & 70.5(6.7) & 73.4(5.5) & 0.72(0.08) \\
      \textbf{Ours} & \textbf{76.3(4.9)} & \textbf{75.8(6.9)} & \textbf{76.7(6.1)} & \textbf{0.78(0.07)} \\
     \hline
    \end{tabular}
    \label{Tab6}
\end{table}

\subsubsection{Effects of JS-Divergence in Reconstruction Task}
We analyzed the effects of the JS-Divergence by varying $\beta \in [0,1]$. The results are shown in \textbf{Table \ref{JS}}. $\beta = 0$ denotes that we only train a subject-similarity discrimination task, achieving 0.72 AUC. As $\beta$ increases, the importance of the JS-Divergence decreases during the model training. $\beta = 1$ denotes that the model is only trained with a square error loss to reconstruct the hidden radiomic features, which achieves 0.70 AUC. In the current study, when $\beta = 0.5$, the prediction performance reached to the peak with a 76.3\% balanced accuracy and a 0.78 AUC.
\begin{table}[ht]
    \centering
    \caption{The effects of the JS-Divergence in the radiomic reconstruction task. $\beta$ indicates a weighting factor of the JS-Divergence in a convex combination in \textbf{Eq \ref{eq1}}. We compared the effects of different $\beta$ for the cognitive deficits risk stratification on the CINEPS dataset by 10-fold cross-validation.}
    \begin{tabular}{ccccc}
     \hline
      & BA (\%) & SEN (\%) & SPE (\%) & AUC\\
      \hline
      $\beta=0.0$&	72.0(5.5)&	70.5(6.7)&	73.4(5.5)&	0.72(0.08) \\
      $\beta=0.2$&	68.3(4.5)&	67.4(5.9)&	69.2(5.3)&	0.69(0.06) \\
      $\beta=0.5$ & \textbf{76.3(4.9)} & \textbf{75.8(6.9)} & \textbf{76.7(6.1)} & \textbf{0.78(0.07)} \\
      $\beta=0.7$ &	71.5(5.1)&	70.5(6.8)&	72.4(6.1)&	0.71(0.06) \\
      $\beta=1.0$ &	69.8(5.1)&	67.2(7.5)&	72.3(6.2)&	0.70(0.07)\\
     \hline
     \label{JS}
    \end{tabular}
\end{table}

\subsubsection{Importance of Subject-Similarity Discrimination Task}
Our proposed method is based on two pretext tasks, i.e., radiomic features reconstruction and subject-similarity discrimination. Depending on the position of masked ROIs, the radiomic features reconstruction task may be sensitive and it is prone to produce mask-dependent feature representations. Therefore, we consider that the subject-similarity discrimination task collaboratively aids the radiomic features reconstruction to learn the mask-invariant features. Two pretext tasks of our method are collaboratively trained to learn the radiomic features, i.e., $ \Tilde{\textbf{f}}_i$. Here, we analyze the importance of the subject-similarity discrimination task in our method. To discover the effects of the subject-similarity discrimination task, we trained our model with different $\lambda$, which indicates the importance of the subject-similarity discrimination task in \textbf{Eq (\ref{eq5})}. The results are shown in $\textbf{Table \ref{Tab8}}$. When $\lambda=0.0$, which denotes that the network is trained only with a radiomic features reconstruction task, achieves 69.8\% on balanced accuracy and 0.70 on AUC. As $\lambda$ increased, the model obtained the best classification performance on cognitive deficits risk classification with 76.3\% on balanced accuracy and 0.78 on AUC. When $\lambda$ continues to increase, the classification performance started to decrease to 71.9\% on balanced accuracy and 0.72 on AUC. Our method achieved the best classification performance on the CINEPS dataset when $\lambda=1.0$. These results further demonstrate the collaborative effectiveness of both the radiomic features reconstruction task and the subject-similarity discrimination task.
\begin{table}[ht]
    \centering
    \caption{The importance of the subject-similarity discrimination task. $\lambda$ indicates a weighting factor of the overall discrimination loss function in $\textbf{Eq (\ref{eq5})}$. We compared the effects of different $\lambda$ for the cognitive deficits risk stratification on the CINEPS dataset by 10-fold cross-validation.}
    \begin{tabular}{ccccc}
     \hline
      & BA (\%) & SEN (\%) & SPE (\%) & AUC\\
      \hline
      $\lambda=0.0$ & 69.8(5.1) & 67.2(7.5) & 72.3(6.2) & 0.70(0.07) \\
      $\lambda=0.5$ & 71.7(5.1)	& 70.5(7.1)	& 72.9(5.9)	& 0.73(0.08) \\
      $\lambda=1.0$ & \textbf{76.3(4.9)} & \textbf{75.8(6.9)} & \textbf{76.7(6.1)} & \textbf{0.78(0.07)} \\
      $\lambda=1.5$ & 74.5(4.6) & 73.5(6.5) & 75.4(5.7) & 0.74(0.06) \\
      $\lambda=2.0$ & 71.9(5.3) & 70.2(7.1) & 73.5(6.5) & 0.72(0.06) \\
     \hline
    \end{tabular}
    \label{Tab8}
\end{table}

\subsubsection{Feature Visualization}
\begin{figure*}[ht]
    \centering
    \includegraphics[width=0.88\textwidth]{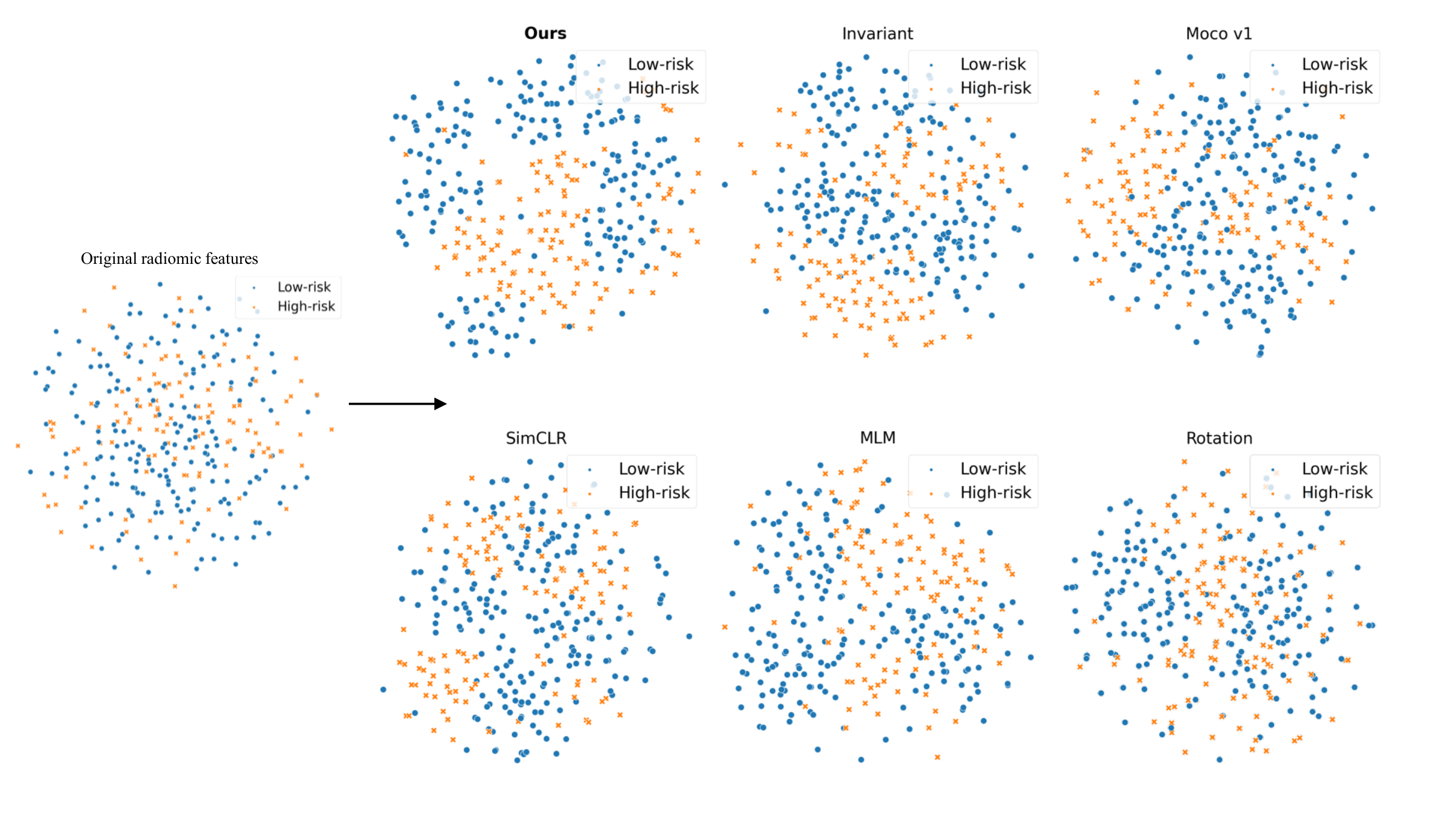}
    \caption{Feature visualization of original radiomic features (left) and self-supervised learned features (right) from the CINEPS dataset. We extracted the features from the last MLP layer of the Transformer and use the T-SNE to visualize the features. Our method showed more separable patterns compared to other self-supervised methods. Orange points represent the low-risk group and blue points represent the high-risk group.}
    \label{tsne}
\end{figure*}

To verify whether our collaborative two pretext tasks can successfully learn the latent radiomic feature representation, we used the T-SNE plot to visualize the learned features after the last MLP layer of our model in \textbf{Figure \ref{tsne}}. The self-supervised learned features from \cite{gidaris2018unsupervised} and \cite{noroozi2016unsupervised} showed a more mixed pattern between positive and negative samples. Compared to other contrastive-based methods, i.e., \cite{chen2020simple, he2020momentum}, Invariant \cite{ye2019unsupervised} had a more separable decision boundary. It is observed that our mode that collaboratively used two pretext tasks showed a clearer potential decision boundary between the two classes. This feature visualization demonstrates that our collaborative self-supervised method can help the model to learn more discriminative patterns.

\subsubsection{Loss Functions Comparison}
\begin{figure}[ht]
    \centering
    \includegraphics[width=8.85cm]{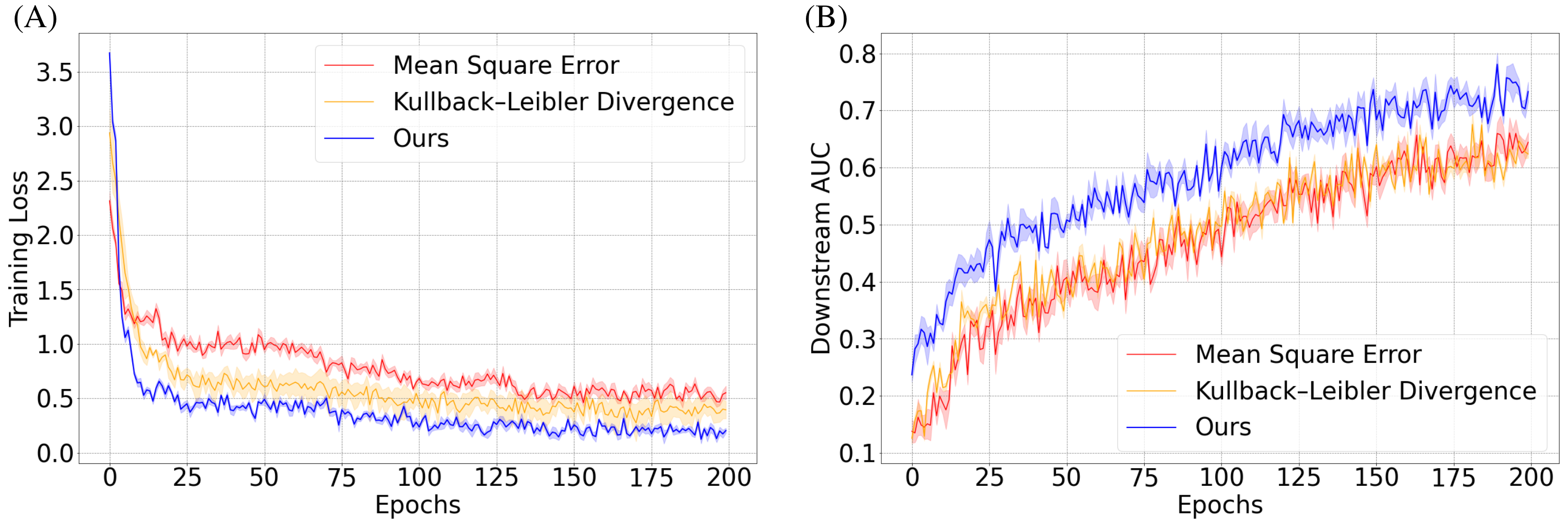}
    \caption{Self-supervised training loss (A) vs. downstream testing AUC (B) on the CINEPS dataset with 10-fold cross-validation. We compare our learning objective loss with MSE and KL-divergence on MLM and SimCLR, respectively.}
    \label{loss}
\end{figure}

Our learning objective loss function can be viewed as a combination of two Bregman divergences, including geometry-based and probabilistic-based. To show the advantage of the loss function of our method, we compared different divergence functions (e.g., MSE and KL-divergence) with our loss function. Note, we used MLM \cite{devlin2018bert} and SimCLR \cite{chen2020simple} as the baseline self-supervised methods and trained models using the MSE and KL-divergence, respectively. To have a fair comparison, we trained each loss function for 200 epochs on the CINEPS dataset with 10-fold cross-validation. As shown in \textbf{Figure \ref{loss}}, we observed our proposed loss function converges faster than MSE and KL-divergence during the pretext task training stage and consistently outperformed MSE and KL-divergence with higher AUC in the downstream classification stage. These results demonstrate the effectiveness of our learning objective loss using the Bregman divergence. 

\subsubsection{Model Performance on Lower Regiment Labeled Data}
\begin{figure}[ht]
    \centering
    \includegraphics[width=8.85cm]{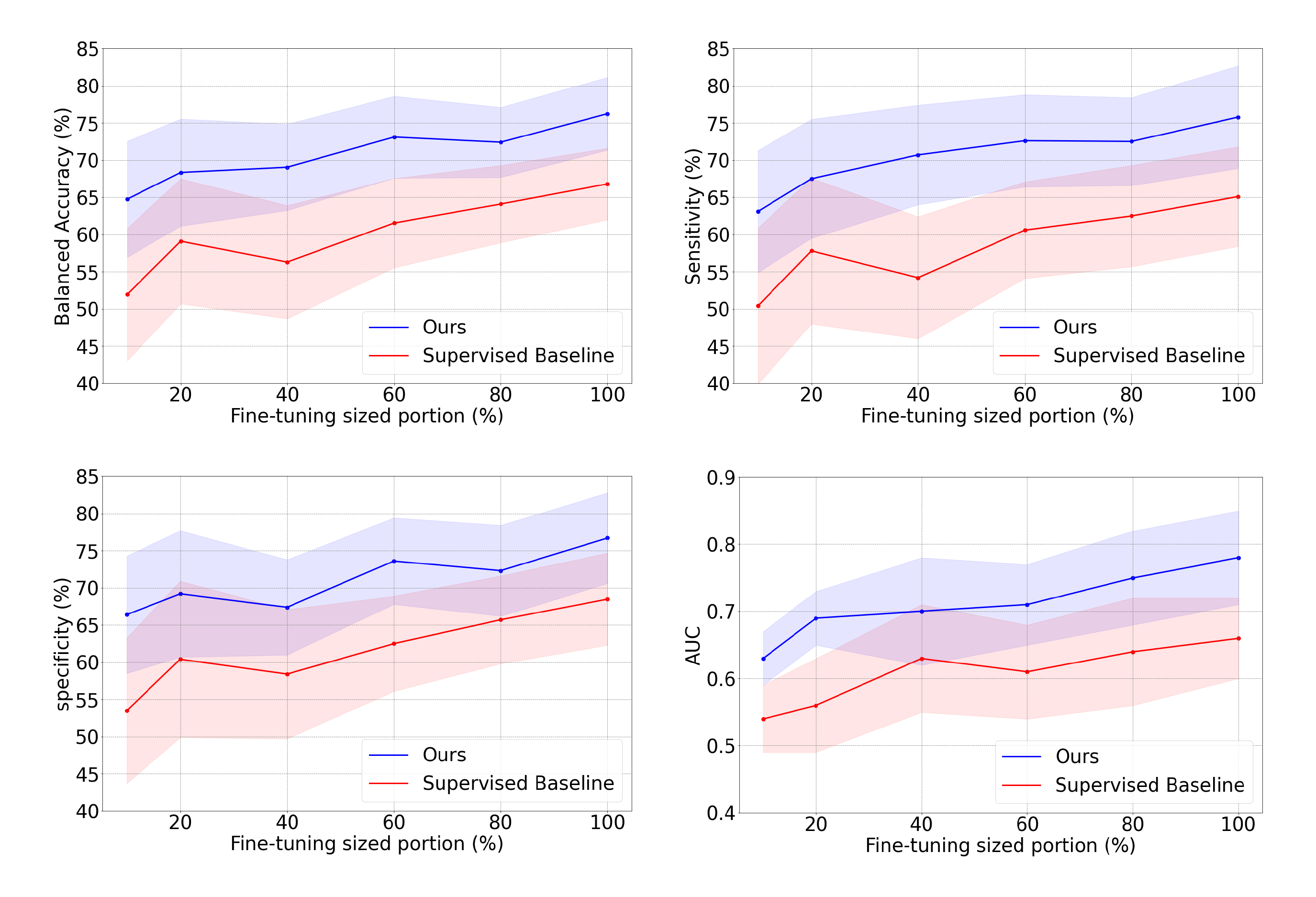}
    \caption{Classification performance on different-sized portions of the CINEPS training dataset. We pretrained our collaborative SSL method using the full CINEPS training dataset and fine-tuned it on the different proportions of the training data.}
    \label{regiment}
\end{figure}

To investigate whether our proposed collaborative SSL method has a robust predictive ability on the lower regiment labeled dataset, we first pretrained our method using the full CINEPS training dataset and then fine-tuned the pretrained model on different-sized portions (i.e., 10\%, 20\%, 40\%, 60\%, and 80\%) of the training dataset. The results are shown in \textbf{Figure \ref{regiment}}. With only 10\% of the training dataset (N=37), the proposed method achieved a 0.63 AUC and 65.0\% balanced accuracy, which was significantly higher ($p<0.001$) than a supervised baseline model with an AUC of 0.54 and a balanced accuracy of 52.0\%. This suggests that the SSL model performed well on the lower regiment labeled dataset. As the sample size increased, the prediction performance of both methods increased, and the proposed collaborative SSL model consistently outperformed the supervised baseline model. These results further demonstrated that our proposed method could retrain competitive prediction performance on small sample size.

\subsubsection{Effects of the Number of Masked ROIs}
In \textbf{Table \ref{masked}}, we showed the model performance with the different number of masked ROIs. As the number of masked ROIs increased, the model performance increased until a critical point (e.g., 30 masked ROIs in our work). We noted that the model performance decreased when the number of masked ROIs was greater than 30. Similar trends were also observed using other competitive SSL models. Note, the purpose of the proposed pretext task is to understand/perceive the input images (i.e., gain prior “knowledge”) through reconstructing the masked ROIs, and such knowledge can thereafter be reused for the downstream task. The observation from \textbf{Table \ref{masked}} may indicate that a small number of the masked ROIs was not challenging enough for the reconstruction task, while a large number was over-challenging. In both scenarios, the pretrained model performed worse than the optimal number we chose in understanding/perceiving the input images.

\begin{table}[ht]
    \centering
    \caption{Effects of the number of masked ROIs in the pretext task on the prediction performance on the CINEPS dataset.  AUC is selected as the evaluation metric.}
    \begin{tabular}{ccccc}
     \hline
      \# ROIs & Rotation \cite{gidaris2018unsupervised} & SimCLR \cite{chen2020simple} & Moco v1 \cite{he2020momentum} & Ours\\ 
      \hline
       10 & 0.59(0.05) & 0.65(0.04) & 0.63(0.04) & 0.73(0.05) \\
       20 &	0.60(0.04) & 0.66(0.08) & 0.67(0.04) &	0.75(0.07) \\
       \textbf{30} &	\textbf{0.64(0.08)} & \textbf{0.68(0.04)} & \textbf{0.69(0.08)} & \textbf{0.78(0.07)} \\
       40 &	0.63(0.07) & 0.65(0.06)	& 0.66(0.07) &	0.76(0.06) \\
     \hline
     \label{masked}
    \end{tabular}
   
\end{table}

\subsubsection{Effects of Network Encoder}
Our method is specifically designed for the MRI radiomic data, which does not have a strong spatial or sequential correlation among ROIs. In this section, we compared different network encoders that have been utilized for computer vision (i.e., CNNs) and natural language processing (i.e., Transformer with position encoding) in \textbf{Table \ref{encoder}}. It is observed that using VGG19 \cite{simonyan2014very} and ResNet18 \cite{he2016deep} as the backbone of the network encoder achieved lower prediction results on a cognitive deficits risk stratification task. This is because the convolutional process of the ResNet18 cannot capture the latent relationship among the radiomic features of various ROIs. A Transformer with position encoding (TransForPos) \cite{vaswani2017attention} achieved higher performance than ResNet18. Notably, our method using a Transformer without a position encoding module significantly outperformed (p<0.001) a VGG19, a ResNet18, and a Transformer with a position encoding module. These results showed the effectiveness of  our network encoder designed specifically for radiomic data.”
\begin{table}[ht]
    \centering
    \caption{Effects of different network encoders for the cognitive deficits risk stratification on the CINEPS dataset}
    \begin{tabular}{ccccc}
     \hline
      & BA (\%) & SEN (\%) & SPE (\%) & AUC\\
      \hline
      VGG19 \cite{simonyan2014very} & 65.3(4.8) & 64.7 (6.2)&	65.9(5.5)&	0.66(0.07)\\
      ResNet18 \cite{he2016deep}& 	64.9(4.6)&	63.5(5.9)&	66.2(5.1)&	0.64(0.08)\\
      TransForPos \cite{vaswani2017attention}&	74.0(5.3)&	73.0(7.5)&	74.9(5.9)&	0.75(0.06)\\
      \textbf{Ours}&	\textbf{76.3(4.9)}&	\textbf{75.8(6.9)}&	\textbf{76.7(6.1)}&	\textbf{0.78(0.07)}\\
     \hline
     \label{encoder}
    \end{tabular}
\end{table}

\subsubsection{Compare with MAE Method}
Our radiomic features reconstruction task is similar to the masked autoencoder (MAE) \cite{he2022masked}, which aims to reconstruct the randomly masked patches from the images. In here, we compared our method with the MAE using an internal validation on the CINEPS dataset and then performed an external validation on the COEPS dataset. We masked several random 2D patches on the radiomic feature map and then designed a pretext task to reconstruct the masked patches. Same as \cite{he2022masked}, we performed random masking of 75\% of the patches on the radiomic features and used the same encoder network (Transformer) for consistency. The results are shown in \textbf{Table \ref{MAE1}} and \textbf{Table \ref{MAE2}}. Compared to the MAE method, our radiomic feature reconstruction task achieved a better prediction performance. It is likely due to the MAE method masks several patches on the radiomic feature map, which does not consider the pathological relationship among the ROIs. These results demonstrated the importance of pathological relationships that existed in ROIs that is helpful for the cognitive deficits risk stratification.
\begin{table}[ht]
    \centering
    \caption{Comparison with the MAE method for the cognitive deficits risks stratification on the CINEPS dataset.}
    \begin{tabular}{ccccc}
     \hline
      & BA (\%) & SEN (\%) & SPE (\%) & AUC\\
      \hline
      MAE \cite{he2022masked} &	68.2(4.5) &	65.5(6.0) &	70.8(5.8) &	0.67(0.04) \\
      Reconstruction &	69.8(5.1) &	67.2(7.5) &	72.3(6.2) &	0.70(0.07) \\
      \textbf{Ours}&	\textbf{76.3(4.9)}&	\textbf{75.8(6.9)}&	\textbf{76.7(6.1)}&	\textbf{0.78(0.07)}\\
     \hline
     \label{MAE1}
    \end{tabular}
\end{table}

\begin{table}[ht]
    \centering
    \caption{Comparison with the MAE method for the cognitive deficits risks stratification on the COEPS dataset.}
    \begin{tabular}{ccccc}
     \hline
      & BA (\%) & SEN (\%) & SPE (\%) & AUC\\
      \hline
      MAE \cite{he2022masked}&	63.9&	60.0&	67.8&	0.62\\
      Reconstruction&	69.8&	70.0&	69.5&	0.65 \\
      \textbf{Ours}&	\textbf{71.4}&	\textbf{70.0}&	\textbf{72.9}&	\textbf{0.70}\\
     \hline
     \label{MAE2}
    \end{tabular}
\end{table}

\subsubsection{Computation Time Comparison Results with CINEPS Dataset}
Computation Time Comparison Results with CINEPS Dataset- In this section, we provided computation time for our method in comparison to SimCLR and Invariant methods. The majority of the time cost of self-supervised learning is pretraining the pretext tasks. So, we only recorded the computation time for one epoch in the pretraining pretext task stage. The results are shown in \textbf{Table \ref{time}}. Our method had the highest computation time compared to the two SimCLR and Invariant, since our method contains two collaborative pretext tasks, while only one pretext task existed in the other two methods. Since the SSL pretraining is commonly applied offline, we expect the extra computational time would not impact future applications of the proposed collaborative SSL method.
\begin{table}[ht]
    \centering
    \caption{Computation time comparison for one epoch during a pretext task pretraining stage on the CINEPS dataset. (Unit: seconds)}
    \begin{tabular}{ccccc}
     \hline
       & SimCLR \cite{chen2020simple} & Invariant \cite{ye2019unsupervised} & \textbf{Ours} \\
       \hline
      Seconds & 587.22 & 547.15 & 874.34 \\
     \hline
     \label{time}
    \end{tabular}
\end{table}

\section{Discussion}
Radiomics techniques are an important tool in assessing various diseases for medical image-based diagnosis, such as neurological impairments  \cite{feng2020mri,bang2021interpretable,park2019radiomics}, cancer \cite{sollini2021pet,ou2020radiomics}, and liver disease \cite{starmans2018classification,jeong2019radiomics,wei2020radiomics}. With the advancement of deep learning techniques (e.g., CNNs), computer-aided diagnosis has shown great promise in supporting the research community. Although deep learning techniques achieve immense success with diagnosis tasks, these methods focus on extracting the deep features from raw images, which usually lose the interpretation for the researchers. Instead of learning directly from raw medical images, radiomics offers an interpretable strategy by extracting high-throughput features to provide statistical descriptions of raw medical images \cite{parmar2015radiomic, park2019radiomics}. However, supervised learning models usually require a large number of human annotations and labeling with domain knowledge, which can be expensive and timely to obtain. SSL has been playing an efficient role to provide solutions for this challenge by learning latent feature representation from data itself without human annotations. 

In this work, we proposed a novel collaborative self-supervised learning method to learn radiomic data. Different than previous SSL methods \cite{gidaris2018unsupervised, noroozi2016unsupervised, he2020momentum, chen2020simple, ye2019unsupervised}, our method is specially designed to learn radiomic data. We designed two collaborative pretext tasks, i.e., radiomic features reconstruction and subject-similarity discrimination, to retrieve latent biological and pathological relations among different ROIs and discover the discriminative patterns from these hidden radiomic features, respectively. Thus, the subject-similarity discrimination task helps the radiomic features reconstruction task to better learn the robust feature representation, which is beneficial for the supervised downstream task, i.e., classification and regression. Our method is validated on three datasets, i.e., simulation, CINEPS, and COEPS datasets, in which our method continued achieving the best prediction performance in terms of both classification and regression tasks among other SSL methods. 

We analyzed our collaborative self-supervised methods using the Bregman divergence. We considered the implicit connection between the learning objective loss function and the Bregman divergence and provided mathematical proof to show the properties of our loss function. As shown in \textbf{Eq (\ref{eq11})}, our loss function considers two perspectives of both geometric divergence and probabilistic divergence during training. To show the advantage of our learning objective function, we conducted a loss function comparison in \textbf{Figure \ref{loss}}, demonstrating the efficiency of training loss convergence and the classification performance of our proposed loss using the Bregman divergence. There are some other pretext tasks, such as rotation \cite{gidaris2018unsupervised}, and jigsaw puzzles \cite{noroozi2016unsupervised}, which learn spatial information from the training data but demonstrated lower performance in this study. This is due to the non-spatial dependencies in radiomic data. Moco v1 \cite{he2020momentum} and SimCLR \cite{chen2020simple} are two self-supervised methods that rely on a larger batch size during training that may not be suitable for the radiomic data, in which the sample size is usually small. MLM \cite{devlin2018bert} predicts masked/hidden features between input and its augmentation but considers less correlation only between augmented data. In our results, Invariant \cite{ye2019unsupervised} also achieved plausible results on our radiomic data, outperforming other hand-crafted predictive-based pretext tasks. Invariant is a contrastive learning method to learn feature embeddings by pulling similar samples and pushing dissimilar samples during the training. The main idea is that the features of the same instance from different data augmentations should be spread-out, i.e., invariant. This ensures that the model can learn a discriminative feature embedding by optimizing the inner products of instance features. With extensive experiments, our method demonstrated the supervisor prediction performance, which also supports the theoretical properties of our method.

Our work has certain limitations. First, we only considered single modality MRI data in the current study. There are some other SSL methods, such as modality-invariant [Li, et al, 2020], which were designed to utilize multimodal medical data to improve model performance. In the future, we will investigate how to apply these methods to learn more representative features for downstream tasks when multimodal data is available. Second, our model is pretrained using CINEPS data containing 362 patients, which is considered a large dataset in the medical imaging domain. The impact of a small size of unlabeled data on pre-training a pretext task is less considered. Whether the proposed self-supervised method with other similar kinds requires a large amount of unlabeled data for pretraining has not been thoroughly investigated and will be of interest to future work. Another limitation is that the external validation dataset (i.e., COEPS dataset) only contained 69 subjects, of which 10 subjects were from the high-risk group. This affects prediction performance, where a few samples can cause a large variation in classifying the high-risk group (i.e., sensitivity measures). Moving forward, we will need to evaluate our method on a larger external dataset to evaluate generalizability.

\section{Conclusion}
In this paper, we proposed a novel collaborative self-supervised learning method for learning radiomic data. Our main idea is to collaboratively train our model on two pretext tasks i.e., a radiomic features reconstruction task and a subject-similarity discrimination task, to learn the latent feature representation from the radiomic data. Extensive experimental results demonstrated the effectiveness of our method. With further refinement, our method may facilitate computer-aided diagnosis applications in clinical practice without large-scale annotated datasets.

\bibliographystyle{IEEEtran}
\bibliography{ref.bib}

\begin{thebibliography}{10}
\providecommand{\url}[1]{#1}
\csname url@samestyle\endcsname
\providecommand{\newblock}{\relax}
\providecommand{\bibinfo}[2]{#2}
\providecommand{\BIBentrySTDinterwordspacing}{\spaceskip=0pt\relax}
\providecommand{\BIBentryALTinterwordstretchfactor}{4}
\providecommand{\BIBentryALTinterwordspacing}{\spaceskip=\fontdimen2\font plus
\BIBentryALTinterwordstretchfactor\fontdimen3\font minus
  \fontdimen4\font\relax}
\providecommand{\BIBforeignlanguage}[2]{{%
\expandafter\ifx\csname l@#1\endcsname\relax
\typeout{** WARNING: IEEEtran.bst: No hyphenation pattern has been}%
\typeout{** loaded for the language `#1'. Using the pattern for}%
\typeout{** the default language instead.}%
\else
\language=\csname l@#1\endcsname
\fi
#2}}
\providecommand{\BIBdecl}{\relax}
\BIBdecl

\bibitem{lambin2012radiomics}
P.~Lambin \emph{et~al.}, ``Radiomics: extracting more information from medical
  images using advanced feature analysis,'' \emph{European journal of cancer},
  vol.~48, no.~4, pp. 441--446, 2012.

\bibitem{ggillies2016radiomics}
R.~GGillies, P.~Kinahan, and H.~Hricak, ``Radiomics: Images are more than
  pictures,'' \emph{They Are Data. Radiology}, vol. 278, no.~2, pp. 563--577,
  2016.

\bibitem{zwanenburg2020image}
A.~Zwanenburg \emph{et~al.}, ``The image biomarker standardization initiative:
  standardized quantitative radiomics for high-throughput image-based
  phenotyping,'' \emph{Radiology}, vol. 295, no.~2, p. 328, 2020.

\bibitem{van2017computational}
J.~J. Van~Griethuysen \emph{et~al.}, ``Computational radiomics system to decode
  the radiographic phenotype,'' \emph{Cancer research}, vol.~77, no.~21, pp.
  e104--e107, 2017.

\bibitem{zwanenburg2019assessing}
A.~Zwanenburg \emph{et~al.}, ``Assessing robustness of radiomic features by
  image perturbation,'' \emph{Scientific reports}, vol.~9, no.~1, pp. 1--10,
  2019.

\bibitem{ou2020radiomics}
X.~Ou \emph{et~al.}, ``Radiomics based on 18f-fdg pet/ct could differentiate
  breast carcinoma from breast lymphoma using machine-learning approach: A
  preliminary study,'' \emph{Cancer medicine}, vol.~9, no.~2, pp. 496--506,
  2020.

\bibitem{isensee2017brain}
F.~Isensee, P.~Kickingereder, W.~Wick, M.~Bendszus, and K.~H. Maier-Hein,
  ``Brain tumor segmentation and radiomics survival prediction: Contribution to
  the brats 2017 challenge,'' in \emph{International MICCAI Brainlesion
  Workshop}.\hskip 1em plus 0.5em minus 0.4em\relax Springer, 2017, pp.
  287--297.

\bibitem{alahmari2018delta}
S.~S. Alahmari, D.~Cherezov, D.~B. Goldgof, L.~O. Hall, R.~J. Gillies, and
  M.~B. Schabath, ``Delta radiomics improves pulmonary nodule malignancy
  prediction in lung cancer screening,'' \emph{Ieee Access}, vol.~6, pp.
  77\,796--77\,806, 2018.

\bibitem{conti2021radiomics}
A.~Conti, A.~Duggento, I.~Indovina, M.~Guerrisi, and N.~Toschi, ``Radiomics in
  breast cancer classification and prediction,'' in \emph{Seminars in cancer
  biology}, vol.~72.\hskip 1em plus 0.5em minus 0.4em\relax Elsevier, 2021, pp.
  238--250.

\bibitem{afshar2019handcrafted}
P.~Afshar, A.~Mohammadi, K.~N. Plataniotis, A.~Oikonomou, and H.~Benali, ``From
  handcrafted to deep-learning-based cancer radiomics: challenges and
  opportunities,'' \emph{IEEE Signal Processing Magazine}, vol.~36, no.~4, pp.
  132--160, 2019.

\bibitem{feng2020mri}
Q.~Feng and Z.~Ding, ``Mri radiomics classification and prediction in
  alzheimer’s disease and mild cognitive impairment: a review,''
  \emph{Current Alzheimer Research}, vol.~17, no.~3, pp. 297--309, 2020.

\bibitem{salvatore2021radiomics}
C.~Salvatore, I.~Castiglioni, and A.~Cerasa, ``Radiomics approach in the
  neurodegenerative brain,'' \emph{Aging Clinical and Experimental Research},
  vol.~33, no.~6, pp. 1709--1711, 2021.

\bibitem{cui2018disease}
L.-B. Cui \emph{et~al.}, ``Disease definition for schizophrenia by functional
  connectivity using radiomics strategy,'' \emph{Schizophrenia bulletin},
  vol.~44, no.~5, pp. 1053--1059, 2018.

\bibitem{park2020differentiating}
Y.~W. Park \emph{et~al.}, ``Differentiating patients with schizophrenia from
  healthy controls by hippocampal subfields using radiomics,''
  \emph{Schizophrenia Research}, vol. 223, pp. 337--344, 2020.

\bibitem{park2020radiomics}
H.~J. Park, B.~Park, and S.~S. Lee, ``Radiomics and deep learning: hepatic
  applications,'' \emph{Korean Journal of Radiology}, vol.~21, no.~4, pp.
  387--401, 2020.

\bibitem{valdora2018rapid}
F.~Valdora, N.~Houssami, F.~Rossi, M.~Calabrese, and A.~S. Tagliafico, ``Rapid
  review: radiomics and breast cancer,'' \emph{Breast cancer research and
  treatment}, vol. 169, no.~2, pp. 217--229, 2018.

\bibitem{he2019machine}
L.~He \emph{et~al.}, ``Machine learning prediction of liver stiffness using
  clinical and t2-weighted mri radiomic data,'' \emph{American Journal of
  Roentgenology}, vol. 213, no.~3, pp. 592--601, 2019.

\bibitem{jeong2019radiomics}
W.~K. Jeong, N.~Jamshidi, E.~R. Felker, S.~S. Raman, and D.~S. Lu, ``Radiomics
  and radiogenomics of primary liver cancers,'' \emph{Clinical and molecular
  hepatology}, vol.~25, no.~1, p.~21, 2019.

\bibitem{beig2020introduction}
N.~Beig, K.~Bera, and P.~Tiwari, ``Introduction to radiomics and radiogenomics
  in neuro-oncology: implications and challenges,'' \emph{Neuro-oncology
  Advances}, vol.~2, no. Supplement\_4, pp. iv3--iv14, 2020.

\bibitem{li2018novel}
H.~Li, N.~A. Parikh, and L.~He, ``A novel transfer learning approach to enhance
  deep neural network classification of brain functional connectomes,''
  \emph{Frontiers in neuroscience}, vol.~12, p. 491, 2018.

\bibitem{lan2019albert}
Z.~Lan, M.~Chen, S.~Goodman, K.~Gimpel, P.~Sharma, and R.~Soricut, ``Albert: A
  lite bert for self-supervised learning of language representations,''
  \emph{arXiv preprint arXiv:1909.11942}, 2019.

\bibitem{chen2020simple}
T.~Chen, S.~Kornblith, M.~Norouzi, and G.~Hinton, ``A simple framework for
  contrastive learning of visual representations,'' in \emph{International
  conference on machine learning}.\hskip 1em plus 0.5em minus 0.4em\relax PMLR,
  2020, pp. 1597--1607.

\bibitem{he2020momentum}
K.~He, H.~Fan, Y.~Wu, S.~Xie, and R.~Girshick, ``Momentum contrast for
  unsupervised visual representation learning,'' in \emph{Proceedings of the
  IEEE/CVF conference on computer vision and pattern recognition}, 2020, pp.
  9729--9738.

\bibitem{kolesnikov2019revisiting}
A.~Kolesnikov, X.~Zhai, and L.~Beyer, ``Revisiting self-supervised visual
  representation learning,'' in \emph{Proceedings of the IEEE/CVF conference on
  computer vision and pattern recognition}, 2019, pp. 1920--1929.

\bibitem{li2020self}
X.~Li, M.~Jia, M.~T. Islam, L.~Yu, and L.~Xing, ``Self-supervised feature
  learning via exploiting multi-modal data for retinal disease diagnosis,''
  \emph{IEEE Transactions on Medical Imaging}, vol.~39, no.~12, pp. 4023--4033,
  2020.

\bibitem{tomar2022self}
D.~Tomar, B.~Bozorgtabar, M.~Lortkipanidze, G.~Vray, M.~S. Rad, and J.-P.
  Thiran, ``Self-supervised generative style transfer for one-shot medical
  image segmentation,'' in \emph{Proceedings of the IEEE/CVF Winter Conference
  on Applications of Computer Vision}, 2022, pp. 1998--2008.

\bibitem{nadif2021unsupervised}
M.~Nadif and F.~Role, ``Unsupervised and self-supervised deep learning
  approaches for biomedical text mining,'' \emph{Briefings in Bioinformatics},
  vol.~22, no.~2, pp. 1592--1603, 2021.

\bibitem{li2021rotation}
X.~Li \emph{et~al.}, ``Rotation-oriented collaborative self-supervised learning
  for retinal disease diagnosis,'' \emph{IEEE Transactions on Medical Imaging},
  vol.~40, no.~9, pp. 2284--2294, 2021.

\bibitem{ji2021does}
S.~Ji, M.~H{\"o}ltt{\"a}, and P.~Marttinen, ``Does the magic of bert apply to
  medical code assignment? a quantitative study,'' \emph{Computers in Biology
  and Medicine}, vol. 139, p. 104998, 2021.

\bibitem{zhuang2019self}
X.~Zhuang, Y.~Li, Y.~Hu, K.~Ma, Y.~Yang, and Y.~Zheng, ``Self-supervised
  feature learning for 3d medical images by playing a rubik’s cube,'' in
  \emph{International Conference on Medical Image Computing and
  Computer-Assisted Intervention}.\hskip 1em plus 0.5em minus 0.4em\relax
  Springer, 2019, pp. 420--428.

\bibitem{bai2019self}
W.~Bai \emph{et~al.}, ``Self-supervised learning for cardiac mr image
  segmentation by anatomical position prediction,'' in \emph{International
  Conference on Medical Image Computing and Computer-Assisted
  Intervention}.\hskip 1em plus 0.5em minus 0.4em\relax Springer, 2019, pp.
  541--549.

\bibitem{feng2019self}
Z.~Feng, C.~Xu, and D.~Tao, ``Self-supervised representation learning by
  rotation feature decoupling,'' in \emph{Proceedings of the IEEE/CVF
  Conference on Computer Vision and Pattern Recognition}, 2019, pp.
  10\,364--10\,374.

\bibitem{peng2017multilevel}
B.~Peng \emph{et~al.}, ``A multilevel-roi-features-based machine learning
  method for detection of morphometric biomarkers in parkinson’s disease,''
  \emph{Neuroscience letters}, vol. 651, pp. 88--94, 2017.

\bibitem{liu2009liver}
P.~Liu, P.~Li, W.~He, and L.-Q. Zhao, ``Liver and spleen volume variations in
  patients with hepatic fibrosis,'' \emph{World Journal of Gastroenterology:
  WJG}, vol.~15, no.~26, p. 3298, 2009.

\bibitem{thompson2020tracking}
D.~K. Thompson \emph{et~al.}, ``Tracking regional brain growth up to age 13 in
  children born term and very preterm,'' \emph{Nature communications}, vol.~11,
  no.~1, pp. 1--11, 2020.

\bibitem{cajanus2019association}
A.~Cajanus \emph{et~al.}, ``The association between distinct frontal brain
  volumes and behavioral symptoms in mild cognitive impairment, alzheimer's
  disease, and frontotemporal dementia,'' \emph{Frontiers in neurology},
  vol.~10, p. 1059, 2019.

\bibitem{bang2021interpretable}
M.~Bang \emph{et~al.}, ``An interpretable multiparametric radiomics model for
  the diagnosis of schizophrenia using magnetic resonance imaging of the corpus
  callosum,'' \emph{Translational psychiatry}, vol.~11, no.~1, pp. 1--8, 2021.

\bibitem{starmans2018classification}
M.~P. Starmans, R.~L. Miclea, S.~R. Van Der~Voort, W.~J. Niessen, M.~G.
  Thomeer, and S.~Klein, ``Classification of malignant and benign liver tumors
  using a radiomics approach,'' in \emph{Medical Imaging 2018: Image
  Processing}, vol. 10574.\hskip 1em plus 0.5em minus 0.4em\relax International
  Society for Optics and Photonics, 2018, p. 105741D.

\bibitem{wei2020radiomics}
J.~Wei \emph{et~al.}, ``Radiomics in liver diseases: Current progress and
  future opportunities,'' \emph{Liver International}, vol.~40, no.~9, pp.
  2050--2063, 2020.

\bibitem{yue2020machine}
H.~Yue \emph{et~al.}, ``Machine learning-based ct radiomics method for
  predicting hospital stay in patients with pneumonia associated with
  sars-cov-2 infection: a multicenter study,'' \emph{Annals of translational
  medicine}, vol.~8, no.~14, 2020.

\bibitem{peng2018distinguishing}
L.~Peng \emph{et~al.}, ``Distinguishing true progression from radionecrosis
  after stereotactic radiation therapy for brain metastases with machine
  learning and radiomics,'' \emph{International Journal of Radiation Oncology*
  Biology* Physics}, vol. 102, no.~4, pp. 1236--1243, 2018.

\bibitem{brunese2020ensemble}
L.~Brunese, F.~Mercaldo, A.~Reginelli, and A.~Santone, ``An ensemble learning
  approach for brain cancer detection exploiting radiomic features,''
  \emph{Computer methods and programs in biomedicine}, vol. 185, p. 105134,
  2020.

\bibitem{gidaris2018unsupervised}
S.~Gidaris, P.~Singh, and N.~Komodakis, ``Unsupervised representation learning
  by predicting image rotations,'' \emph{arXiv preprint arXiv:1803.07728},
  2018.

\bibitem{noroozi2016unsupervised}
M.~Noroozi and P.~Favaro, ``Unsupervised learning of visual representations by
  solving jigsaw puzzles,'' in \emph{European conference on computer
  vision}.\hskip 1em plus 0.5em minus 0.4em\relax Springer, 2016, pp. 69--84.

\bibitem{spitzer2018improving}
H.~Spitzer, K.~Kiwitz, K.~Amunts, S.~Harmeling, and T.~Dickscheid, ``Improving
  cytoarchitectonic segmentation of human brain areas with self-supervised
  siamese networks,'' in \emph{International Conference on Medical Image
  Computing and Computer-Assisted Intervention}.\hskip 1em plus 0.5em minus
  0.4em\relax Springer, 2018, pp. 663--671.

\bibitem{liu2021self}
X.~Liu \emph{et~al.}, ``Self-supervised learning: Generative or contrastive,''
  \emph{IEEE Transactions on Knowledge and Data Engineering}, 2021.

\bibitem{goodfellow2016deep}
I.~Goodfellow, Y.~Bengio, and A.~Courville, \emph{Deep learning}.\hskip 1em
  plus 0.5em minus 0.4em\relax MIT press, 2016.

\bibitem{wu2020self}
D.~Wu, H.~Ren, and Q.~Li, ``Self-supervised dynamic ct perfusion image
  denoising with deep neural networks,'' \emph{IEEE Transactions on Radiation
  and Plasma Medical Sciences}, vol.~5, no.~3, pp. 350--361, 2020.

\bibitem{chen2019self}
L.~Chen, P.~Bentley, K.~Mori, K.~Misawa, M.~Fujiwara, and D.~Rueckert,
  ``Self-supervised learning for medical image analysis using image context
  restoration,'' \emph{Medical image analysis}, vol.~58, p. 101539, 2019.

\bibitem{ye2019unsupervised}
M.~Ye, X.~Zhang, P.~C. Yuen, and S.-F. Chang, ``Unsupervised embedding learning
  via invariant and spreading instance feature,'' in \emph{Proceedings of the
  IEEE/CVF Conference on Computer Vision and Pattern Recognition}, 2019, pp.
  6210--6219.

\bibitem{grill2020bootstrap}
J.-B. Grill \emph{et~al.}, ``Bootstrap your own latent-a new approach to
  self-supervised learning,'' \emph{Advances in Neural Information Processing
  Systems}, vol.~33, pp. 21\,271--21\,284, 2020.

\bibitem{vaswani2017attention}
A.~Vaswani \emph{et~al.}, ``Attention is all you need,'' \emph{Advances in
  neural information processing systems}, vol.~30, 2017.

\bibitem{he2022masked}
K.~He, X.~Chen, S.~Xie, Y.~Li, P.~Doll{\'a}r, and R.~Girshick, ``Masked
  autoencoders are scalable vision learners,'' in \emph{Proceedings of the
  IEEE/CVF Conference on Computer Vision and Pattern Recognition}, 2022, pp.
  16\,000--16\,009.

\bibitem{saito1994new}
T.~Saito and J.-I. Toriwaki, ``New algorithms for euclidean distance
  transformation of an n-dimensional digitized picture with applications,''
  \emph{Pattern recognition}, vol.~27, no.~11, pp. 1551--1565, 1994.

\bibitem{wang2005euclidean}
L.~Wang, Y.~Zhang, and J.~Feng, ``On the euclidean distance of images,''
  \emph{IEEE transactions on pattern analysis and machine intelligence},
  vol.~27, no.~8, pp. 1334--1339, 2005.

\bibitem{bai2013graph}
L.~Bai and E.~R. Hancock, ``Graph kernels from the jensen-shannon divergence,''
  \emph{Journal of mathematical imaging and vision}, vol.~47, no.~1, pp.
  60--69, 2013.

\bibitem{lin1991divergence}
J.~Lin, ``Divergence measures based on the shannon entropy,'' \emph{IEEE
  Transactions on Information theory}, vol.~37, no.~1, pp. 145--151, 1991.

\bibitem{kullback1951information}
S.~Kullback and R.~A. Leibler, ``On information and sufficiency,'' \emph{The
  annals of mathematical statistics}, vol.~22, no.~1, pp. 79--86, 1951.

\bibitem{bregman1967relaxation}
L.~M. Bregman, ``The relaxation method of finding the common point of convex
  sets and its application to the solution of problems in convex programming,''
  \emph{USSR computational mathematics and mathematical physics}, vol.~7,
  no.~3, pp. 200--217, 1967.

\bibitem{corso2021challenge}
F.~Corso \emph{et~al.}, ``The challenge of choosing the best classification
  method in radiomic analyses: Recommendations and applications to lung cancer
  ct images,'' \emph{Cancers}, vol.~13, no.~12, p. 3088, 2021.

\bibitem{parikh2021perinatal}
N.~A. Parikh \emph{et~al.}, ``Perinatal risk and protective factors in the
  development of diffuse white matter abnormality on term-equivalent age
  magnetic resonance imaging in infants born very preterm,'' \emph{The Journal
  of Pediatrics}, vol. 233, pp. 58--65, 2021.

\bibitem{bayley2006bayley}
N.~Bayley, \emph{Bayley scales of infant and toddler development}.\hskip 1em
  plus 0.5em minus 0.4em\relax PsychCorp, Pearson, 2006.

\bibitem{gousias2012magnetic}
I.~S. Gousias \emph{et~al.}, ``Magnetic resonance imaging of the newborn brain:
  manual segmentation of labelled atlases in term-born and preterm infants,''
  \emph{Neuroimage}, vol.~62, no.~3, pp. 1499--1509, 2012.

\bibitem{makropoulos2018developing}
A.~Makropoulos \emph{et~al.}, ``The developing human connectome project: A
  minimal processing pipeline for neonatal cortical surface reconstruction,''
  \emph{Neuroimage}, vol. 173, pp. 88--112, 2018.

\bibitem{devlin2018bert}
J.~Devlin, M.-W. Chang, K.~Lee, and K.~Toutanova, ``Bert: Pre-training of deep
  bidirectional transformers for language understanding,'' \emph{arXiv preprint
  arXiv:1810.04805}, 2018.

\bibitem{simonyan2014very}
K.~Simonyan and A.~Zisserman, ``Very deep convolutional networks for
  large-scale image recognition,'' \emph{arXiv preprint arXiv:1409.1556}, 2014.

\bibitem{he2016deep}
K.~He, X.~Zhang, S.~Ren, and J.~Sun, ``Deep residual learning for image
  recognition,'' in \emph{Proceedings of the IEEE conference on computer vision
  and pattern recognition}, 2016, pp. 770--778.

\bibitem{park2019radiomics}
Y.~W. Park \emph{et~al.}, ``Radiomics and machine learning may accurately
  predict the grade and histological subtype in meningiomas using conventional
  and diffusion tensor imaging,'' \emph{European radiology}, vol.~29, no.~8,
  pp. 4068--4076, 2019.

\bibitem{sollini2021pet}
M.~Sollini \emph{et~al.}, ``Pet/ct radiomics in breast cancer: Mind the step,''
  \emph{Methods}, vol. 188, pp. 122--132, 2021.

\bibitem{parmar2015radiomic}
C.~Parmar, P.~Grossmann, D.~Rietveld, M.~M. Rietbergen, P.~Lambin, and H.~J.
  Aerts, ``Radiomic machine-learning classifiers for prognostic biomarkers of
  head and neck cancer,'' \emph{Frontiers in oncology}, vol.~5, p. 272, 2015.

\end{thebibliography}
\end{document}